\documentclass[10pt,journal,compsoc]{IEEEtran}
\usepackage{amssymb,amsmath,amsthm}
\usepackage[lined,boxed,commentsnumbered,ruled]{algorithm2e}
\usepackage{algorithmic} 
\usepackage{graphicx}
\usepackage{stmaryrd}
\SetSymbolFont{stmry}{bold}{U}{stmry}{m}{n}
\usepackage{multirow,url}
\usepackage{color,cite}
\usepackage{cite}
\usepackage{tikz,pgf}
\usetikzlibrary{fadings,shapes,arrows,shadows}
\usepackage{makecell}
\usepackage{array}
\usepackage{overpic} 
\usepackage{subfigure}
\usepackage{url}

\newtheorem{proposition}{Proposition}
\newtheorem{definition}{Definition}

\newtheorem{mechanism}{Mechanism}

\theoremstyle{plain}

\def\x{\boldsymbol{x}}
\def\z{\boldsymbol{z}}
\def\p{\boldsymbol{p}}
\def\b{\boldsymbol{b}}
\def\c{\boldsymbol{c}}
\def\bv{\boldsymbol{u}}
\def\v{\boldsymbol{v}}
\def\pv{\boldsymbol{q}}

\def\wx{\boldsymbol{\widetilde{x}}}
\def\wz{\boldsymbol{\widetilde{z}}}
\def\wp{\boldsymbol{\widetilde{p}}}
\def\wpv{\boldsymbol{\widetilde{q}}}

\def\I{\mathcal{I}}
\def\J{\mathcal{J}}
\def\D{\mathcal{D}}
\def\K{\mathcal{K}}

\def\T{\mathcal{T}}

\def\S{\mathcal{S}}

\def\A{\mathcal{A}}

\def\R{\mathcal{R}}

\def\d{\mathrm{d}}

\def\eq{\triangleq}

\def\mech{\Omega}
\def\wmech{\widetilde{\mech}}

\begin{document}

\title{Data-Centric Mobile Crowdsensing}

\author{
        Changkun~Jiang, 
        Lin~Gao,~\IEEEmembership{Senior~Member,~IEEE,}
        Lingjie~Duan,~\IEEEmembership{Member,~IEEE,}
        and~Jianwei~Huang,~\IEEEmembership{Fellow,~IEEE}
\IEEEcompsocitemizethanks{
\IEEEcompsocthanksitem Part of the results have appeared in IEEE GLOBECOM, Dec. 2016 \cite{JGDH}.
\IEEEcompsocthanksitem Changkun Jiang and Jianwei Huang are with the Network Communications and Economics Lab, Department of Information Engineering, The Chinese University of Hong Kong, Shatin, N.T., Hong Kong, China. \protect\\
E-mail: \{jc012, jwhuang\}@ie.cuhk.edu.hk
\IEEEcompsocthanksitem Lin Gao is with the School of Electronic Information and Engineering, Harbin Institute of Technology (Shenzhen),  Shenzhen, China. \protect\\
E-mail: gaolin@hitsz.edu.cn
\IEEEcompsocthanksitem Lingjie Duan is with the Engineering Systems and Design Pillar, Singapore University of Technology and Design, 8 Somapah Road, Singapore. \protect\\
E-mail: lingjie\_duan@sutd.edu.sg}
}

\IEEEtitleabstractindextext{%
\begin{abstract}
Mobile crowdsensing (MCS) is a promising sensing paradigm that leverages the diverse embedded sensors in massive mobile devices.
A key objective in MCS is to efficiently schedule mobile users to perform multiple sensing tasks.
Prior work mainly focused on interactions between the task-layer and the user-layer, without considering tasks' similar data requirements and users' heterogeneous sensing capabilities.
In this work, we propose a three-layer data-centric MCS model by introducing a new data-layer between tasks and users, enable different tasks to reuse the same data items, hence effectively leverage both task similarities and user heterogeneities. 
We formulate a joint task selection and user scheduling problem based on the new framework, aiming at maximizing social welfare. We first analyze the centralized optimization problem with the statistical information of tasks and users, and show the bound of the social welfare gain due to data reuse. Then we consider the two-sided information asymmetry of selfish task-owners and users, and propose a decentralized market mechanism for achieving the centralized social optimality. In particular, considering the NP-hardness of the optimization, we propose a truthful two-sided randomized auction mechanism that ensures computational efficiency and a close-to-optimal performance. Simulations verify the effectiveness of our proposed model and mechanism.
\end{abstract}

}

\addtolength{\abovedisplayskip}{-1mm}
\addtolength{\belowdisplayskip}{-1mm}

\maketitle

\IEEEdisplaynontitleabstractindextext

\IEEEpeerreviewmaketitle

\IEEEraisesectionheading{\section{Introduction}\label{sec:introduction}}
\subsection{Background and Motivations}
\IEEEPARstart{T}{he} proliferation of hand-held mobile devices with rich
embedded sensors has enabled a new sensing paradigm known as \emph{Mobile CrowdSensing (MCS)} \cite{MCSSurvey}, where individual mobile users are involved in performing the sensing tasks by using their mobile devices. Due to the low deploying cost and the high sensing coverage, this new sensing paradigm has attracted a broad range of applications such as urban dynamic mining, public safety, and environment monitoring
\cite{waze,opensignal,UrbanAtmosphere}.
In a general multi-task MCS system (e.g., PRISM \cite{Platform1} and Medusa~\cite{Platform2}),  each sensing task is first initiated and announced by a task planner (task owner) via a web portal. Then the task is assigned to a pool of mobile users (registered in the system), who will perform the sensing task accordingly (e.g.,
sensing the required data and sending the collected data to the system).
While performing a sensing task, mobile users consume their own device resources such as battery energy and CPU time, hence incur certain costs.
Thus, users may not be willing to participate in MCS, unless they receive proper rewards to compensate their costs.

Many prior studies (e.g., \cite{SmartphoneCollaboration,DYang,HJin,TieLuo,YanminZhu}) have studied the problem of incentivizing users to participate in the MCS system. These works focused on the interactions of tasks and users (e.g., the assignment of tasks among
users through a proper  matching), without considering the common data
requirements (hence the potential data reuse) among multiple tasks
and the heterogeneous sensing capabilities of different users.
In a practical system, however, there is a high likelihood that multiple tasks require some common data \cite{MCSSurvey}.
For example, the road traffic data at a particular time and location may be useful for Waze\footnote{Waze (\url{www.waze.com}) is the world's largest community-based traffic and navigation app.}, Uber, and Google Traffic simultaneously.
Therefore, it is likely to cause duplicated data sensing and processing in a multi-task scenario, if multiple tasks are completed separately by the same user.
Moreover, in a practical system, users may have different sensing capabilities due to factors such as locations and device types. For example, it is easier for a user to sense the data close to her current location.
Thus, it is more flexible and efficient to schedule users on the data level than on the task level.

To complete multiple tasks more efficiently, it is critical to identify the common data requirements of these tasks and enable the reuse of sensory data across different tasks.
Specifically, some practical MCS platforms (e.g., PRISM \cite{Platform1} and Medusa~\cite{Platform2})
have allowed task developers to specify their data requirements in a high-level language. Then, they identify and reuse the common data across multiple tasks in order to reduce or avoid duplicated sensing and processing.
There are several advantages enable data reuse in the MCS system.
First, data is digital goods and can be reused without additional cost.
Second, multiple tasks can share a large pool of mobile users collectively through the platform.
Third, by reusing data across different tasks, the overall system efficiency can be improved.
A similar MCS architecture has been discussed in \cite{MCSSurvey} as a future vision. However, \cite{MCSSurvey} did not provide any theoretical framework or analysis about the performance gain that can be achieved through data reuse.

\subsection{Novelty and Contributions}

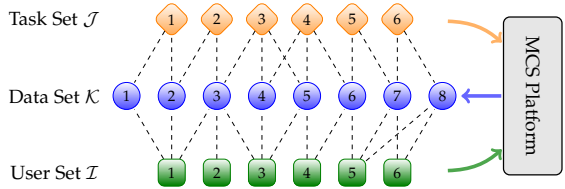
\begin{figure}[t]
\centering
\def\layersep{1.7cm}
\centering
\scalebox{0.6}{
\begin{tikzpicture}[shorten >=1pt,->,draw=black!50, node distance=\layersep]
    \tikzstyle{data}=[circle,fill=black!25,minimum size=17pt,inner sep=0pt]
    \tikzstyle{task}=[diamond,rounded corners,fill=black!25,minimum size=23pt,inner sep=0pt]
    \tikzstyle{user}=[rectangle,rounded corners,fill=black!25,minimum size=17pt,inner sep=0pt]
    \tikzstyle{Task}=[draw,task,orange, top color= white,bottom color=orange!70,text=black];
    \tikzstyle{User}=[draw,user,green!50!black!100, top color= white,bottom color=green!50!black!100,text=black];
    \tikzstyle{Data}=[draw,data,blue, top color= white,bottom color=blue!70,text=black];

    \tikzstyle{Platform}=[rectangle, rounded corners, thin, thick,
                               black, fill= gray!20, draw, text=black,
                               minimum width=0.6cm, minimum height=3.5cm];

    \tikzstyle{annot} = [text width=4.2em, text centered]
    \tikzstyle{annot1} = [text width=3.2em, text centered]

    \foreach \name / \x in {1,...,6}
        \node[Task] (I-\name) at (\x,0) {\x};

    \foreach \name / \x in {1,...,8}
        \path[xshift=-1cm]
            node[Data] (H-\name) at (\x,-\layersep) {\x};

    \foreach \name / \x in {1,...,6}
            \path[xshift=0cm]
                node[User] (O-\name) at (\x,-2*\layersep) {\x};

   \node[] (I-7) at (7,0) {};
   \node[] (H-9) at (7.3,-\layersep) {};
   \node[] (O-7) at (7,-2*\layersep) {};

    \foreach \source in {1}
        \foreach \dest in {1,2}
            \path[black,dashed,-] (I-\source) edge (H-\dest);
    \foreach \source in {2}
        \foreach \dest in {2,3}
            \path[black,dashed,-] (I-\source) edge (H-\dest);
    \foreach \source in {3}
        \foreach \dest in {3,4,5}
            \path[black,dashed,-] (I-\source) edge (H-\dest);
    \foreach \source in {4}
        \foreach \dest in {4,5,6}
            \path[black,dashed,-] (I-\source) edge (H-\dest);
    \foreach \source in {5}
        \foreach \dest in {6,7}
            \path[black,dashed,-] (I-\source) edge (H-\dest);
    \foreach \source in {6}
        \foreach \dest in {7,8}
            \path[black,dashed,-] (I-\source) edge (H-\dest);

    \foreach \source in {1}
        \foreach \dest in {1,2,3}
            \path[black,dashed,-] (O-\source) edge (H-\dest);
    \foreach \source in {2}
        \foreach \dest in {3}
            \path[black,dashed,-] (O-\source) edge (H-\dest);
    \foreach \source in {3}
        \foreach \dest in {3,4,5}
            \path[black,dashed,-] (O-\source) edge (H-\dest);
    \foreach \source in {4}
        \foreach \dest in {5,6}
            \path[black,dashed,-] (O-\source) edge (H-\dest);
    \foreach \source in {5}
        \foreach \dest in {6,7,8}
            \path[black,dashed,-] (O-\source) edge (H-\dest);
    \foreach \source in {6}
        \foreach \dest in {7,8}
            \path[black,dashed,-] (O-\source) edge (H-\dest);

    \node[left of=H-1, node distance=1.6cm] (hl) {{\large Data Set $\mathcal{K}$}};
    \node[above of=hl]  (h2) {\large Task Set $\mathcal{J}$};
    \node[below of=hl] (h3) {{\large User Set $\mathcal{I}$}};


    \node[Platform,annot1] at (9,-1.7) (PF) {};
    \node[annot] at (PF){\rotatebox{-90}{\large MCS Platform}};

    \path[draw=orange!60,line width=0.8mm,->] (I-7) edge[bend  left=20] (PF);
    \draw[blue!60,line width=0.8mm,<-] (H-9) -- (PF);
    \path[draw=green!50!black!70,line width=0.8mm,->] (O-7) edge[bend  right=20] (PF);
\end{tikzpicture}
}

\caption{Three-layer data-centric mobile crowdsensing model. The MCS platform acts as  the social planner to maximize the total social welfare.
}
\label{Fig:SystemModel}
\end{figure}

In this work, we propose a novel three-layer \emph{data-centric} MCS model, consisting of a data layer, a task layer, and a user layer,
which is different from the traditional two-layer \emph{task-centric} model in \cite{SmartphoneCollaboration,DYang,HJin,TieLuo,YanminZhu} (with the task layer and the user layer only).
Specifically, in our data-centric model, tasks and users are connected through the data layer, that is, \emph{each task is translated to a set of data items that it requires, and each user is connected to a set of data items that she can sense.
Moreover, different tasks may require a common data item  (hence can reuse the data item reported by users), and different users may be able to sense the same data item (hence compete with each other for the sensing opportunity).}
Thus, it is able to leverage both the \emph{task similarity} (in terms of data requirements) and the \emph{user heterogeneity} (in terms of sensing capabilities).
Fig.~\ref{Fig:SystemModel} illustrates such a crowdsensing model with 6 tasks, 6 users, and 8 data items, where task $1$ requires data items $\{1,2\}$, and user $1$ is able to sense data items $\{1,2,3\}$ simultaneously.

In such a  data-centric model, the MCS platform (social planner) collects the  data requirements of tasks and the  sensing capabilities of users, and then decides
whether and how to complete these tasks by a proper set of users efficiently.
Formally,
\begin{itemize}
  \item Which tasks can be completed?
  \item Which users will be scheduled for sensing which data?
\end{itemize}
We focus on the optimal task selection and user scheduling
that maximize the social welfare, where the social welfare is the difference between the total values of completed tasks and the total costs of scheduled users.
We are interested in understanding two key questions. The first question is what is the performance gain due to data reuse? Such a gain depends on the numbers of tasks and users  as well as their data requirements. We want to analytically derive the performance gain for any given sets of tasks and users.
Solving this problem is very challenging, as it is NP-hard due to the combinatorial nature.
Moreover, it requires the complete information regarding the task values and users' sensing costs, which are often private information of task owners
and users, respectively. 
Hence the second related question is how to achieve the optimal performance gain in a practical scenario with a limited computational capability and incomplete information? One approach is to design a \emph{truthful} incentive mechanism to elicit such private information from both task owners and users.
However, some well-known truthful incentive mechanisms such as the standard VCG auction \cite{vcg} are not suitable for our problem due to the high computational complexity.

To answer the above two questions, we first conduct a statistical analysis for the bound of the performance gain due to data reuse, and show that such a gain can be quite significant. To reach the optimal performance bound, a social planner needs to make a centralized decision on behalf of all task owners and users. However, in practice, task owners and users are selfish and unwilling to report their private information about task values and sensing cost, which makes the centralized implementation infeasible. To address this issue, we will design an incentive mechanism that satisfies the individual rationality and incentive compatibility for all task owners and users. Such a mechanism also needs to have a low computational complexity and ensures a proper budget balance. To satisfy all the above requirements, we propose a two-sided randomized auction that is tractable for theoretical performance analysis. 

Specifically, we resort to the \emph{randomized auction} framework \cite{ra} for our mechanism design, with the MCS platform acting as the \emph{auctioneer} and the participating task owners and users acting as the \emph{bidders}. 
We propose a truthful randomized auction, consisting of  (i) a randomized allocation rule, which picks up an ``allocation'' (i.e., a feasible solution to the task selection and user scheduling) randomly from a set of feasible solutions according to some probability distribution, and (ii) a payment rule, which assigns a payment for each task owner and user under the chosen allocation. 
Randomized auctions have been adopted for the resource allocation in wireless networking \cite{RAuction}, covering problems \cite{Swamy}, cloud computing~\cite{Zongpeng}, and electricity markets \cite{Zongpeng2}.
The key difference between our randomized auction and those in \cite{RAuction,Swamy,Zongpeng,Zongpeng2} is that our auction is \emph{two-sided}, i.e., we need to decide both the task selection (task values) and the user scheduling (sensing costs) under \emph{mutual information asymmetry}; while the auction models in \cite{RAuction,Swamy,Zongpeng,Zongpeng2} are single-sided (i.e., considering either values or costs), hence are not directly applied to our problem setting. 

\begin{table*}
\renewcommand{\arraystretch}{1.3}
\caption{Key Results In This Paper}\label{tab:keyresults}
\centering
\begin{tabular}{|c|c|c|c|c|c|}
\hline
\textbf{Mechanism} & \textbf{Truthfulness} & \textbf{Efficiency} & \textbf{Budget Balance} & \textbf{Complexity} & \textbf{Section no.}\\
\hline
1: Two-sided VCG auction & Truthful & Optimal & No & NP-hard &\ref{subsec:VCG}\\
\hline
2: Fractional VCG auction & Truthful & Fractional optimal & No & Polynomial &\ref{subsec:auction}\\
\hline
3: Randomized auction & Truthful in expectation & Close-to-optimal & No & Polynomial &\ref{subsec:auction}\\
\hline
4: Randomized auction (reserve price)& Truthful in expectation & Close-to-optimal & Yes & Polynomial &\ref{reservepriceauction}\\
\hline
\end{tabular}
\end{table*}

The proposed randomized auction is truthful (in expectation), in the sense that task owners and users have no incentives to misreport their private task values and sensing costs, respectively.
We further show that the proposed truthful randomized auction is computationally efficient, as both the allocation rule and payment rule can be computed in polynomial time. In summary, we list the key results and the corresponding section numbers in Table \ref{tab:keyresults}.
Our main results and key contributions are summarized as follows.
\begin{itemize}
\item \emph{Novel Data-Centric Crowdsensing Model:} To our best knowledge, this is the first work that analytically exploits data reuse across multiple tasks and analyzes the performance bound due to data reuse in MCS.
We propose a novel three-layer data-centric model, which leverages both the task similarity and the user sensing heterogeneity.

\item \emph{Performance Bound Analysis of Data Reuse:} Our analytical result  shows that the lower bound of the social welfare gain due to data reuse is 2 for a single reusable data item, when the number of tasks and the number of users are comparable to each other. That is, the social welfare is at least doubled by exploiting data reuse across tasks for a single reusable data item.
As the number of data items increases, the social welfare gain due to data reuse also increases.

\item \emph{Randomized Auction Mechanism for Incomplete Information and Limited Computation:} To address the complexity issue of the joint task selection and user scheduling and elicit the two-sided private information from both task owners and users, we propose a  truthful two-sided randomized auction mechanism, which is computationally efficient, individually rational, and incentive compatible (truthful) in expectation. We further design a randomized auction mechanism with a reserve price to achieve the budget balance, with a slightly reduced social welfare.

\item \emph{Observations and Insights:}
Simulations show that (i) the social welfare gain due to data reuse increases with the task  similarity and reaches up to $1300\%$ in our simulations, and (ii) our proposed randomized auction achieves at least $90\%$ of the maximum social welfare.
Furthermore, the increase of the task similarity increases the social welfare with data reuse, as the required number of users performing the tasks can be reduced. However, the increase of the task similarity decreases the social welfare without data reuse, due to the increased user competition.
\end{itemize}

The rest of the paper is organized as follows.
In Section~\ref{sec:systemmodel}, we present the system model.
In Section~\ref{sec:performancebound}, we theoretically analyze the performance bound.
In Section~\ref{sec:vcg}, we first propose the two-sided auction framework to address the incomplete information problem.
Then we characterize the randomized auction mechanism. 
To make the randomized auction budget-balanced, we further propose a reserve price based randomized auction in Section~\ref{sec:extension}.
We present the simulation results in Section \ref{sec:simulation}, and conclude in Section \ref{sec:conclusion}.

\section{System Model}\label{sec:systemmodel}
In this section, we first present the crowdsensing platform model, task model, and user model.  Then we formulate the social welfare maximization problem.

\subsection{Crowdsensing Platform Model}
We consider a general multi-task MCS platform consisting of
a set $\J=\{1,2,\cdots,J\}$ of tasks, a set $\I = \{1,2,\cdots,I\}$ of mobile users, and a set $\K=\{1,2,\cdots,K\}$ of target data items.
Each data item $k \in \K$ is characterized by a set of fine-grained parameters such as the data type, location, and time.\footnote{For example, a data item can be the temperature of a room at 11 am every day, the traffic speed of a highway at 6 pm, or a raw sensor reading such as GPS and light sensor.}
Each task $j \in \J $ is associated with a set of data requirements $\K_j \subseteq \K $, and each user $i \in \I$ is able to sense a specific set $\S_i \subseteq \K $ of data items.
As different tasks can reuse the same data item, there may exist two tasks $j_1$ and $j_2$ with overlapping data requirements, i.e., $\K_{j_1}  \bigcap  \K_{j_2} \neq \emptyset$.
Fig.~\ref{Fig:SystemModel} illustrates such a three-layer data-centric MCS model.

The crowdsensing model operates in a time-slotted manner. We divide the whole time period into multiple \emph{time slots}, where each time slot can be an hour or a day, depending on the data precisions of tasks or users.
At the beginning of each time slot,
(i) each task owner registers her task on the platform, indicating the data requirements of the task and the potential value that she can achieve when the task is completed;
and
(ii) each user reports her information on the platform, indicating the sensing capability of the user (i.e., the set of data items that she can sense) and the potential cost for sensing any subset of data items within her capability.
After collecting the reported information from all task owners and users, the platform decides the task selection (i.e., selecting a   set of tasks to be completed) and the user scheduling (i.e., scheduling a   set of users to sense the associated data items of the selected tasks).

\subsection{Task Model}

Recall that each task $j \in \J $ is associated with a set of data requirements $\K_j \subseteq \K $ in the time slot that we focus on, and a task value $v_j > 0$ when it is completed. The task value $v_j$ is the \emph{private information} of task $j$, and cannot be observed by the platform, users, or other tasks. This is one of the two key challenges for optimizing a crowdsensing system with data reuse.
We assume that a task $j$ is completed if and only if each of its required data items in $\K_j$ has been sensed by at least one user.
Let $z_j \in \{0, 1\}$ denote whether a task $j \in
\J$ is completed, and $y_k \in \{0, 1\}$ denote whether a data item $k \in \K$ is sensed by at least one user. Then, for each task $j\in\J$, we have the following constraint:
\begin{equation}\label{eq:zj}
z_j  \leq  y_k,\quad \forall k \in \K_j.
\end{equation}

Given a feasible task selection $\boldsymbol{z} = (z_j,j\in\J)$, the total achieved value (of all completed tasks) is:
\begin{equation}\label{eq:Vz}
V(\boldsymbol{z}) = \sum_{j\in \J}  v_j \cdot z_j.
\end{equation}

\subsection{User Model}
Recall that each user $i \in \I $ is able to sense a set $\S_i  $ of data items in the time slot that we focus on.
The platform can schedule user $i$ to sense a subset $\S \subseteq \S_i $ of data items within her sensing capability, associated with a sensing cost $c_i(\S)$.
Let $x_i(\S) \in \{0, 1\} $  denote whether a user $i$ is scheduled to sense a data set $\S \subseteq \S_i $.
When $\S = \emptyset$, then $x_i(\emptyset) = 1$ denotes that user $i$ is not scheduled to sense any data set, hence has a zero sensing cost, i.e., $c_i( \emptyset ) = 0$.

We assume that a user can only be scheduled to sense one data set within her capability in the target time slot. That is, for each user $i\in \I$, we have the following \emph{user scheduling constraint}:
\begin{equation}\label{eq:xi}
\sum_{\S \subseteq \S_i  } x_i(\S) = 1.
\end{equation}
If a user is scheduled to sense multiple data sets, say $\S_i^{1} $ and $ \S_i^{2}$, it is equivalent to scheduling the user to sense the data set $\S_i^{1}\bigcup\S_i^{2}$.
Let $\boldsymbol{x}_i = (x_i(\S), \ \S \subseteq\S_i )$ denote the scheduling vector of user $i$.
Then, given a feasible user scheduling $\boldsymbol{x} = (\boldsymbol{x}_i,i\in\I)$, the total incurred sensing cost (of all scheduled users) is:
\begin{equation}\label{eq:Cx}
C(\boldsymbol{x}) = \sum_{i\in \I} \sum_{\S \subseteq \S_i }  c_i (\S) \cdot x_i (\S).
\end{equation}

Let $y_{ki} \in \{0, 1\}$ denote whether a data item $k$ is sensed by a user $i$, that is,
$
y_{k i} =  \sum_{\S \subseteq \S_i :k\in \S} x_i (\S).
$
Recall that $y_k \in \{0, 1\}$ denotes whether a data item $k \in \K$ is sensed by at least one user. Then, for each data item $k\in \K$,  we have the following  constraint:
\begin{equation}\label{eq:yk}
y_k \leq \sum_{i \in \I} y_{k i}.
\end{equation}

Moreover, we denote $\boldsymbol{c}_i = (c_i(\S), \ \S \subseteq\S_i )$ as the sensing cost vector of user $i$ for all possible subsets of data items that she can sense.
In practice, the sensing cost vector $\boldsymbol{c}_i$ is the \emph{private information} of user $i$, and cannot be observed by the platform, task owners, or other users. This is the second key challenge for optimizing a crowdsensing system with data reuse. Besides the task values and the user sensing costs, all the other information (i.e., the data requirement $\K_j$ of task $j$ and the sensing capability $\S_i$ of user $i$) are \emph{public information} to the MCS platform. This is because both task owners and users need to first register with the MCS platform, and have no incentives to misreport the information.\footnote{For task owner $j\in\J$, under-reporting the data requirement $\K_j$ means that her data will never be completed (which leads to a task value of 0) , and over-reporting $\K_j$ causes additional cost for achieving the same task value. For user $i\in\I$, under-reporting the sensing capability $\S_i$ weakens her own competitiveness, and over-reporting $\S_j$ can be easily detected by the MCS platform.}

\subsection{Social Welfare Maximization}

The social welfare $W(\boldsymbol{x}, \boldsymbol{z})$ is defined as the difference between the
total value $V(\boldsymbol{z})$ of all completed tasks and the total
  sensing cost $C(\boldsymbol{x})$ of all scheduled users, i.e.,
\begin{equation}\label{eq:Sxz}
W(\boldsymbol{x}, \boldsymbol{z}) = V(\boldsymbol{z}) - C(\boldsymbol{x}).
\end{equation}
The objective of the platform is to decide the best task selection $\boldsymbol{z}$ and   user scheduling $\boldsymbol{x}$ to maximize the social welfare $W(\boldsymbol{x}, \boldsymbol{z})$.
Formally, we can formulate such a joint task selection and user scheduling problem (P1) as follows.
\begin{align}\label{eq:p1}
\mbox{P1:} & & \max_{\boldsymbol{x}, \boldsymbol{y}, \boldsymbol{z}} & ~~~ V(\boldsymbol{z}) - C(\boldsymbol{x})
\notag\\
& & \mbox{s.t.} & ~~~  \eqref{eq:zj}\mbox{--}\eqref{eq:yk}, \quad \forall i \in \I, j \in \J, k \in \K;
\notag\\
& & \mbox{var.} & ~~~  x_i(\S) \in \{0, 1\},\quad \forall \S\in\S_i , i \in \I ;
\notag\\
& & & ~~~   z_j \in \{0, 1\},\quad \forall j \in \J ;\notag\\
& &  & ~~~ y_k \in \{0, 1\},\quad \forall   k \in \K.\notag
\end{align}
Here $\boldsymbol{y} \eq (y_k, k\in\K)$ is an intermediate variable denoting whether each data item is sensed (by at least one user), which bridges the relationship   between the task selection and the user scheduling.
It is easy to see that Problem P1 is a binary integer linear programming problem.
Let $\{\boldsymbol{x}^o, \boldsymbol{y}^o, \boldsymbol{z}^o\}$ denote the optimal solution to P1.
For presentation clarity, we will also write $\{\boldsymbol{x}^o, \boldsymbol{y}^o, \boldsymbol{z}^o\}$ as $\{\boldsymbol{x}^o (\boldsymbol{c},\boldsymbol{v}), \boldsymbol{y}^o(\boldsymbol{c},\boldsymbol{v}), \boldsymbol{z}^o(\boldsymbol{c},\boldsymbol{v})\}$, as all of them are functions of the user sensing costs $\boldsymbol{c} = (\boldsymbol{c}_i, i\in\I)$ and the task values $\boldsymbol{v}=(v_j,j\in\J)$.

There are two main issues that we are interested in investigating. First, what is the performance gain via data reuse?\footnote{Due to the page limit, we put the problem formulation for the social welfare maximization \emph{without} data reuse to  Appendix A.}
Second, how to achieve the above performance gain in a practical scenario with a limited computational capability and incomplete information? Solving Problem P1 is very challenging. Problem P1 is NP-hard (as the special case of P1 can be reduced to the set cover problem),
and hence it is important to design a low-complexity approximate algorithm to find an approximate solution.
Meanwhile, solving Problem P1 requires the complete system information including the data requirements and values of all tasks as well as the sensing capabilities and costs of all users. However, as we have mentioned earlier, users' sensing costs and tasks' values are their private information, and cannot be observed by the MCS platform.
Thus, we need to design a truthful incentive mechanism to elicit such private information.

To this end, we will first study the performance gain of data reuse and analyze the performance bound in Section \ref{sec:performancebound}, where the social planner makes decisions for all users and task owners. Then we will focus on incentive mechanisms design to address the complexity and incomplete information issues in Section \ref{sec:vcg}.

\section{Performance Bound Analysis of Data Reuse}\label{sec:performancebound}
In this section, we analyze the performance bound with data reuse. We start with the simplest case of one data item, the analysis of which provides us insights into the more general case. We will consider multiple tasks and multiple users, with explicitly closed results derived for the case of two tasks and two users. Then we will consider the more general case of multiple data items, multiple users, and multiple tasks through numerical studies.

\subsection{Order Statistics Basics}
The analysis in Section \ref{sec:performancebound} will rely on the tools from Order Statistics \cite{OrderStatistics}, the basics of which will be reviewed in this subsection. 

Let $X_1, X_2, \cdots,X_n$ be $n$ random variables sampled from a continuous distribution with the p.d.f. $f(x)$ and the c.d.f. $F(x)$. The corresponding order statistics are the sequence arranged in the nondecreasing order. The smallest of the sample is denoted by $X_{1:n}$, i.e., $X_{1:n}=\min(X_1, X_2, \cdots,X_n)$, the $m$-th smallest of the sample is denoted by $X_{m:n}$, and finally the largest of the sample is denoted by $X_{n:n}$, i.e., $X_{n:n}=\max(X_1, X_2, \cdots,X_n)$. Then we have $X_{1:n}\leq\cdots \leq X_{m:n}\leq\cdots\leq X_{n:n}$.
The p.d.f. of $X_{m:n}$ for $1\leq m\leq n$ is
\begin{equation}
f_{m:n}(x)=n\binom{n-1}{m-1}(F(x))^{m-1}(1-F(x))^{n-m}f(x).
\end{equation}

Now we derive the joint distribution of all $n$ order statistics and the joint distribution of the first $s$ $(1\leq s\leq n)$ order statistics, respectively. First notice that if $F(x)$ is continuous, then with probability 1 the order statistics of the samples take \emph{distinct} values. Hence it is reasonable to assume the realizations of the $n$ order statistics $X_{1:n}\leq X_{2:n}\leq\cdots\leq X_{n:n}$ to be $x_{1:n} <x_{2:n} <\cdots< x_{n:n}$. This means that the original random variables $X_1,X_2,\cdots,X_n$ are restrained to take on the values $x_{m:n} (m = 1,2,\cdots, n)$, which by symmetry assigns the equal probability for each of the $n!$ permutations of $( 1, 2,\cdots, n)$. Hence, we have the joint density function of all $n$ order statistics to be 
\begin{equation}
f_{1,2,\cdots,n:n}(x_1,x_2,\cdots ,x_n)= n!\prod\limits_{m=1}^{n}f(x_m), x_1<\cdots <x_n.
\end{equation}
Furthermore, for the first $s$ $(1\leq s\leq n)$ order statistics $X_{1:n}\leq\cdots\leq X_{s:n}$, by symmetry we need to assign the equal probability for each of the $n!$ permutations of $( 1, 2,\cdots, n)$, and then take into account the first $s$, i.e., $(1,2,\cdots,s)$. Hence, we have the joint density function of the first $s$ $(1\leq s\leq n)$ order statistics to be 
\begin{equation}\label{eq:jointdistribution}
f_{1,2,\cdots,s:n}(x_1,x_2,\cdots ,x_s)= n!\prod\limits_{m=1}^{s}f(x_m), x_1<\cdots <x_s.
\end{equation}
Similarly, if we define the nonincreasing order statistics as $X_{1:n}\geq X_{2:n}\geq \cdots\geq X_{n:n}$, then the joint distribution of the first $s$ $(1\leq s\leq n)$ nonincreasing order statistics is 
\begin{equation}
f_{1,2,\cdots,s:n}(x_1,x_2,\cdots ,x_s)= n!\prod\limits_{m=1}^{s}f(x_m), x_1>\cdots >x_s.
\end{equation}

Given the above preliminaries, we will conduct the performance bound analysis in the following subsections.

\subsection{Analysis for a Single Reusable Data Item}
In the case with a single data item, each task requires the data item to be completed, and each user can sense the same data item. We assume that the task values $(v_j, j\in\J)$ follow the i.i.d. distribution with the same p.d.f. $f(v)$, and the user costs $ (\boldsymbol{c}_i, i\in\I)$ follow the i.i.d. distribution with the same p.d.f. $g(c)$. 
\subsubsection{Analysis without Data Reuse}
In the scenario without data reuse, since all tasks require the same data, user $i$ has the same cost $c_i$ to complete any of the tasks. We propose the following method to analyze this scenario. We sort the task values by the descending order, i.e., $v_{1:J}\geq v_{2:J}\geq \cdots \geq v_{J:J}$, and sort sensing costs by the ascending order, i.e., $c_{1:I}\leq c_{2:I}\leq \cdots \leq c_{I:I}$. Then, there is a threshold $m$ such that the $m$-th task value is no greater than the $m$-th user cost. The social welfare maximization selection selects tasks with values $v_{1:J},\cdots, v_{m:J}$ and users with sensing costs $c_{1:I},\cdots, c_{m:I}$. Hence, we have $\min\{I,J\}+1$ cases in terms of the threshold $m$ as follows.
\begin{itemize}
\item Case 0: $v_{1:J}<c_{1:I}$, then the task and user selection set is $\T_0=\{(v,c):v_{1:J}<c_{1:I}\}$;
\item Case $m, 1\leq m\leq \min\{I,J\}-1$: $v_{m:J}\geq c_{m:I}$ and $v_{m+1:J}< c_{m+1:I}$, then the task and user selection set is $\T_m=\{(v,c):v_{m:J}\geq c_{m:I}, v_{m+1:J}< c_{m+1:I}\}$;
\item Case $\min\{I,J\}$: $v_{\min\{I,J\}:J}\geq c_{\min\{I,J\}:I}$, then the task and user selection set is $\T_{\min\{I,J\}}=\{(v,c):v_{\min\{I,J\}:J}\geq c_{\min\{I,J\}:I}\}$.
\end{itemize}

For case 0, no tasks or users will be selected, and the social welfare is 0.

For case $m,1\leq m\leq \min\{I,J\}-1$,  tasks with values $v_{1:J},\cdots, v_{m:J}$ and users with costs  $c_{1:I},\cdots,  c_{m:I}$ will be selected.  Hence, the social welfare $SW_{n}[m]$ is
$$ 
SW_{n}[m]=\int_{\T_m}\sum\limits_{k=1}^{m}\left(v_{k:J}-c_{k:I}\right)f_{1,\cdots, m+1:J}g_{1,\cdots, m+1:I}\mathrm{d}v\mathrm{d}c.
$$

For case $\min\{I,J\}$,  tasks with values $v_{1:J},\cdots, v_{\min\{I,J\}:J}$ and users with costs  $c_{1:I},\cdots, c_{\min\{I,J\}:I}$ will be selected.  Hence, the social welfare $SW_{n}[\min\{I,J\}]$ is
$$ 
\begin{aligned}
SW_{n}[\min\{I,J\}]&=\int_{\T_{\min\{I,J\}}}\sum\limits_{k=1}^{\min\{I,J\}}\left(v_{k:J}-c_{k:I}\right)\times\\ & ~~~~~~~~~~~~~f_{1,\cdots,\min\{I,J\}:J}g_{1,\cdots,\min\{I,J\}:I}\mathrm{d}v\mathrm{d}c.
\end{aligned}
$$

Hence, the total social welfare without data reuse is the sum of the social welfare in the  $\min\{I,J\}+1$ cases, i.e.,
\begin{align}
& SW_{n}=\sum_{m=1}^{\min\{I,J\}-1}SW_{n}[m]+SW_{n}[\min\{I,J\}].
\end{align}

By transforming the domains of integration $\T_m (m=1,\cdots,\min\{I,J\})$, it turns out that we can derive the social welfare without data reuse in the following explicit form. That is,
\begin{align}
& SW_{n} 
=\sum_{m=1}^{\min\{I,J\}}\int_{0}^{1}\int_{0}^{v_{1:J}}\cdots\int_{0}^{v_{m-1:J}}\cdots\int_{0}^{v_{\min\{I,J\}-1:J}}\notag\\
&~~\int_{0}^{v_{m:J}}\int_{c_{1:I}}^{v_{2:J}}\cdots\int_{c_{m-1:I}}^{v_{m:J}}\int_{c_{m:I}}^{1}\int_{c_{m+1:I}}^{1}\cdots\int_{c_{\min\{I,J\}-1:I}}^{1}\notag\\
&~~(v_{m:J}-c_{m:I})[(\min\{I,J\})!]^2\prod_{k=1}^{\min\{I,J\}}f(v_k)g(c_k)\notag\\
&~~\d c_{\min\{I,J\}:I}\cdots\d c_{2:I}\d c_{1:I}\d v_{\min\{I,J\}:J}\cdots\d v_{2:J}\d v_{1:J}.\label{eq:SOnoreuse}
\end{align}
In particular, if  ($v_j,j\in\J)$ and $(c_i,i\in\I)$ follow i.i.d. uniform distributions, the integral can be further simplified since $\prod_{k=1}^{\min\{I,J\}}f(v_k)g(c_k)=1$. Under the assumption of uniform distributions, we have the following results in Proposition \ref{prop:withoutdatareuse}.
\begin{proposition}[Social Welfare without Data Reuse]\label{prop:withoutdatareuse}
Under the i.i.d. uniform distributions of  ($v_j,j\in\J)$ and $(c_i,i\in\I)$, the social welfare without data reuse is $J/4$ when the number of users and the number of tasks are identically and sufficiently large, i.e., $I=J\to\infty$. The social welfare without data reuse is $J/2$ when $ J$ is finite and  $I\to\infty$, and is $I/2$ when $I$ is finite and  $J\to\infty$.
\end{proposition}

\subsubsection{Analysis with Data Reuse}
In the scenario with data reuse, we have two possible cases:
\begin{itemize}
\item Case I: If $\min\{c_i,i\in\I\}\leq \sum_{j\in\J}v_j$, then the minimum cost user will sense the data of all tasks, and all tasks will be selected. Let $c=\min\{c_i,i\in\I\}$ and $v=\sum_{j\in\J}v_j$, then the task and user selection set is $\R=\{(v,c):\min\{c_i\}\leq \sum_{j\in\J}v_j\}$.  Let the p.d.f. of $\min\{c_i,i\in\I\}$ be $g_{\min\{c_i,i\in\I\}}(c)$ and the p.d.f. of $\sum_{j\in\J}v_j$ be $f_{\sum_{j\in\J}v_j}(v)$, then the social welfare is
$$
\int_{\R}(v-c)f_{\sum_{j\in\J}v_j}(v)g_{\min\{c_i,i\in\I\}}(c)\mathrm{d}v\mathrm{d}c.
$$
\item Case II: If $\min\{c_i,i\in\I\}> \sum_{j\in\J}v_j$, then no task or user will be selected. The social welfare is 0.
\end{itemize}
Hence, the total social welfare with data reuse is the sum of the social welfare in the two cases, which is given by
\begin{align}
SW_{r}=\int_{\R}(v-c)f_{\sum_{j\in\J}v_j}(v)g_{\min\{c_i,i\in\I\}}(c)\mathrm{d}v\mathrm{d}c.
\end{align}

By transforming the domain of integration $\R$, it turns out that we can derive the social welfare with data reuse in the following explicit form. That is,
\begin{align}
& SW_{r} 
=\int_{0}^{1}\int_{0}^{v}(v-c)f_{\sum_{j\in\J}^{}v_j}(v)g_{\min\{c_i,i\in\I\}}(c)\mathrm{d}c\mathrm{d}v\notag\\
&+\sum_{j=1}^{J-1}\int_{j}^{j+1}\!\!\!\int_{0}^{1}(v-c)f_{\sum_{j\in\J}^{}v_j}(v)g_{\min\{c_i,i\in\I\}}(c)\mathrm{d}c\mathrm{d}v.\label{eq:SOreuse}
\end{align}
In particular, if  ($v_j,j\in\J)$ and $(c_i,i\in\I)$ follow i.i.d. uniform distributions, then the term $\sum_{j\in\J}^{}v_j$ follows the Irwin-Hall distribution with the p.d.f.
$$
\textstyle f_{\sum_{j\in\J}^{}v_j}(v)=\frac{1}{2(J-1)!}\sum_{j=0}^{J}(-1)^j\binom{J}{j}(v-j)^{n-1}\mathrm{sgn}(v-j),
$$
where $\mathrm{sgn}(\cdot)$ is the sign function, i.e., $\mathrm{sgn}(x)=-1$ if $x<0$, $\mathrm{sgn}(x)=0$ if $x=0$, and $\mathrm{sgn}(x)=1$ if $x>0$. The term $\min\{c_i,i\in\I\}$ follows a distribution with the following p.d.f.,
$$
g_{\min\{c_i,i\in\I\}}(c)=I(1-c)^{I-1}.
$$
We have the following result with uniform distribution.
\begin{proposition}[Social Welfare with Data Reuse]\label{prop:withdatareuse}
Under the i.i.d. uniform distributions of  ($v_j,j\in\J)$ and $(c_i,i\in\I)$, the social welfare with data reuse is given by
\begin{equation}
SW_r=\frac{J}{2}-1+\frac{I}{I+1}.
\end{equation}
That is, the social welfare with data reuse is $J/2$ when the number of users $I\to \infty$.
\end{proposition}

\subsubsection{Performance Bound}
We show the performance bound by comparing the social welfare with and without data reuse. In particular, we define the (relative) performance gain due to data reuse as
\begin{equation}\label{eq:performancebound}
\gamma=\frac{SW_{r}}{SW_{n}}.
\end{equation}

Based on Propositions \ref{prop:withoutdatareuse} and \ref{prop:withdatareuse}, we have the following results on the relative performance gain $\gamma$ defined in (\ref{eq:performancebound}). 
\begin{proposition}[Performance Bound]\label{prop:SocialWelfareGain}
Under the i.i.d. uniform distributions of  ($v_j,j\in\J)$ and $(c_i,i\in\I)$, 
\begin{itemize}
\item when the numbers of users and tasks are identical and sufficiently large, e.g., $I=J\to\infty$, the lower bound of the relative performance gain is
$\gamma_{lower~bound}=(J/2)/(J/4)=2.$
That is, the social welfare is at least doubled by exploiting data reuse across tasks;
\item when the number of users is sufficiently large, e.g., $I\to\infty$, with a limited $J$, the lower bound of the relative performance gain is $\gamma_{lower~bound}=(J/2)/(J/2)=1.$ That is, the social welfare due to data reuse is at least the same as that without data reuse;  
\item when the number of tasks is sufficiently large, e.g., $J\to\infty$, with a limited $I$, the lower bound of the relative performance gain is $\gamma_{lower~bound}=(J/2)/(I/2)=J/I.$ That is, the social welfare due to data reuse is much larger than that without data reuse.
\end{itemize}
\end{proposition}
 
  \begin{figure*}
  \begin{minipage}[t]{0.32\textwidth}
  \centering
  \includegraphics[scale=0.32]{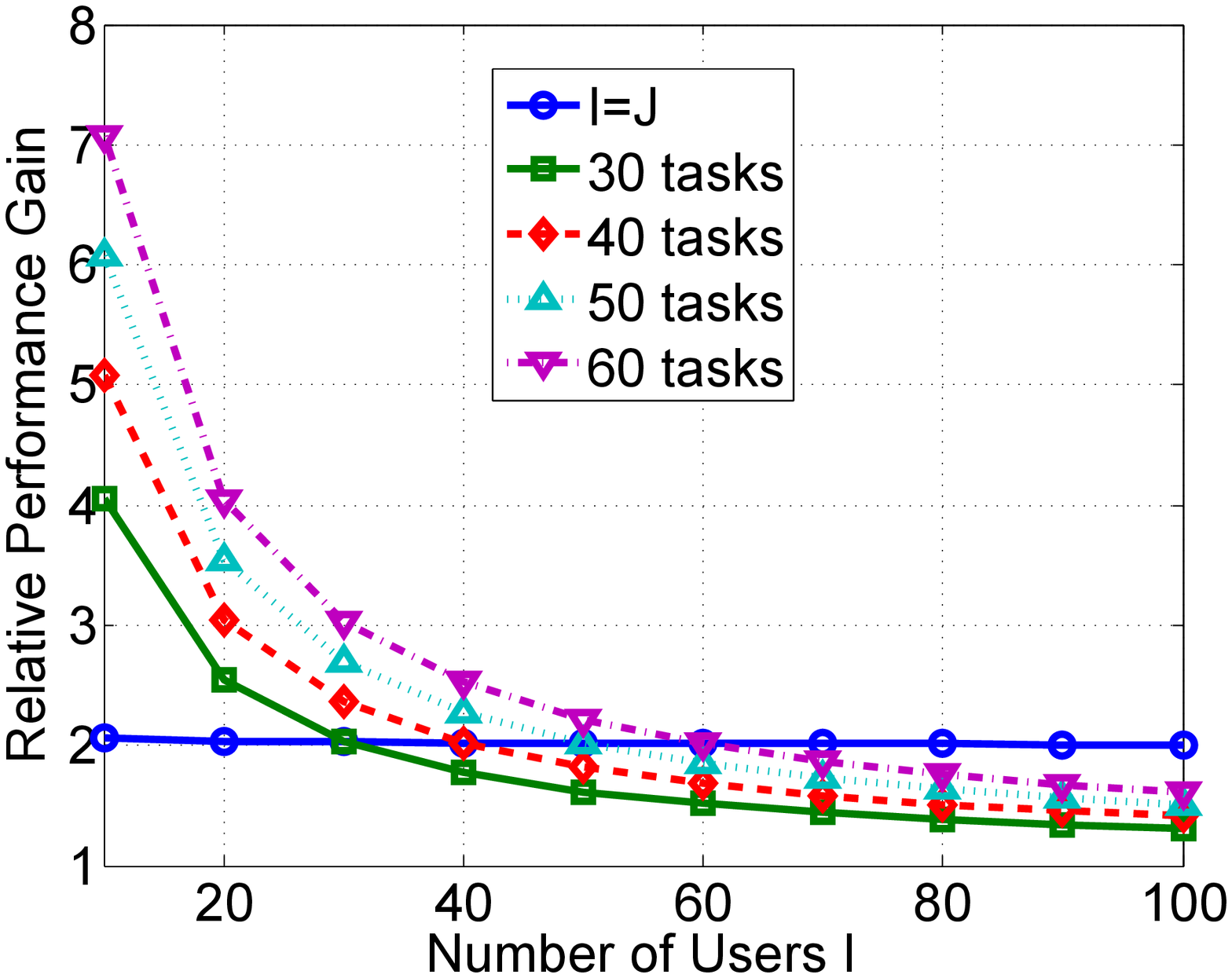}
  \caption{Impact of the number of users on the relative performance gain}\label{Fig:MonteCarlo}
  \vspace{-4mm}
  \end{minipage}%
  \begin{minipage}[t]{0.02\textwidth}
  ~~~
  \end{minipage}%
  \begin{minipage}[t]{0.32\textwidth}
  \centering
  \includegraphics[scale=0.32]{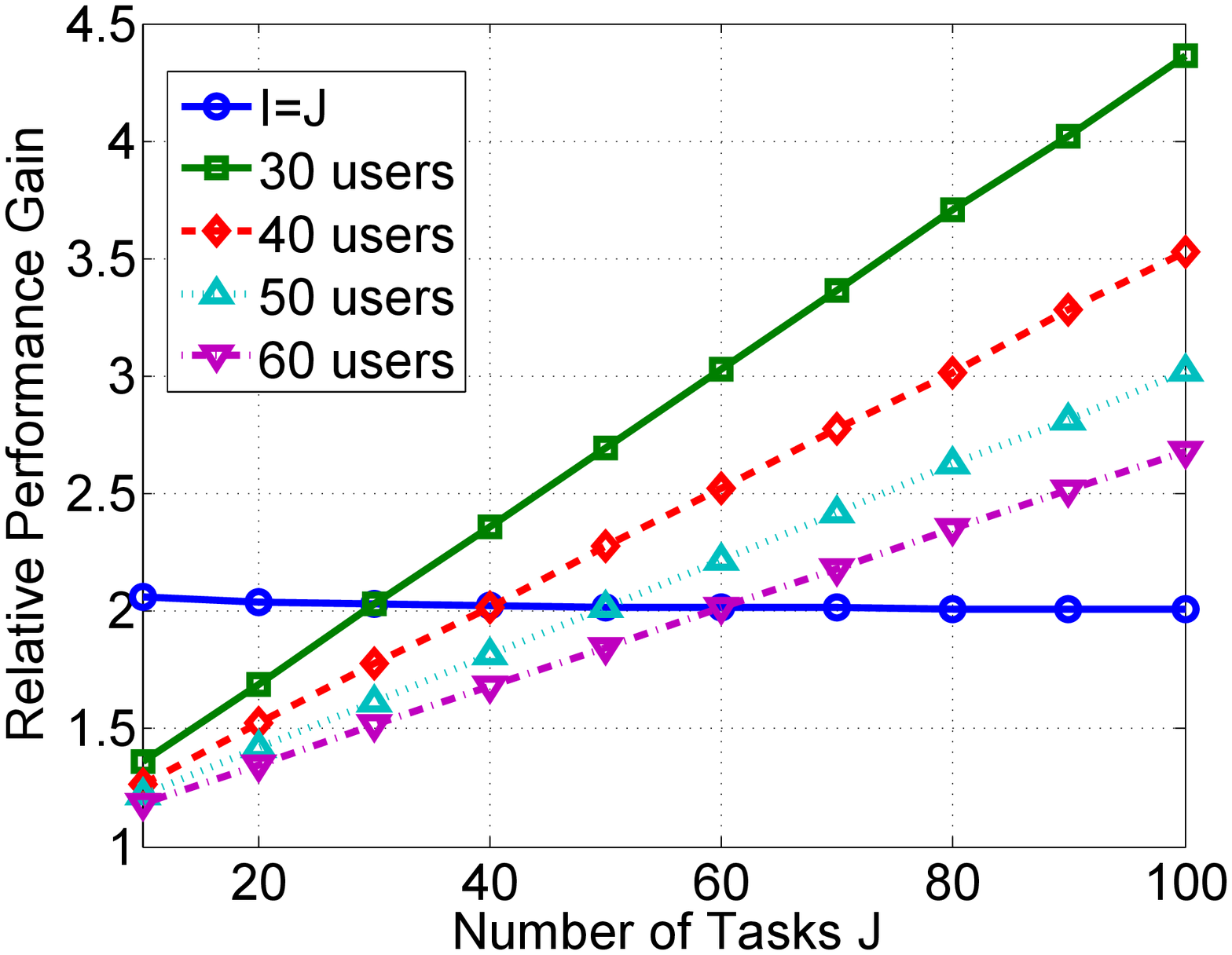}
  \caption{Impact of the number of tasks on the relative performance gain}\label{Fig:MonteCarlo2}
  \vspace{-4mm}
  \end{minipage}%
  \begin{minipage}[t]{0.02\textwidth}
  ~~~
  \end{minipage}%
  \begin{minipage}[t]{0.32\textwidth}
  \centering
  \includegraphics[scale=0.32]{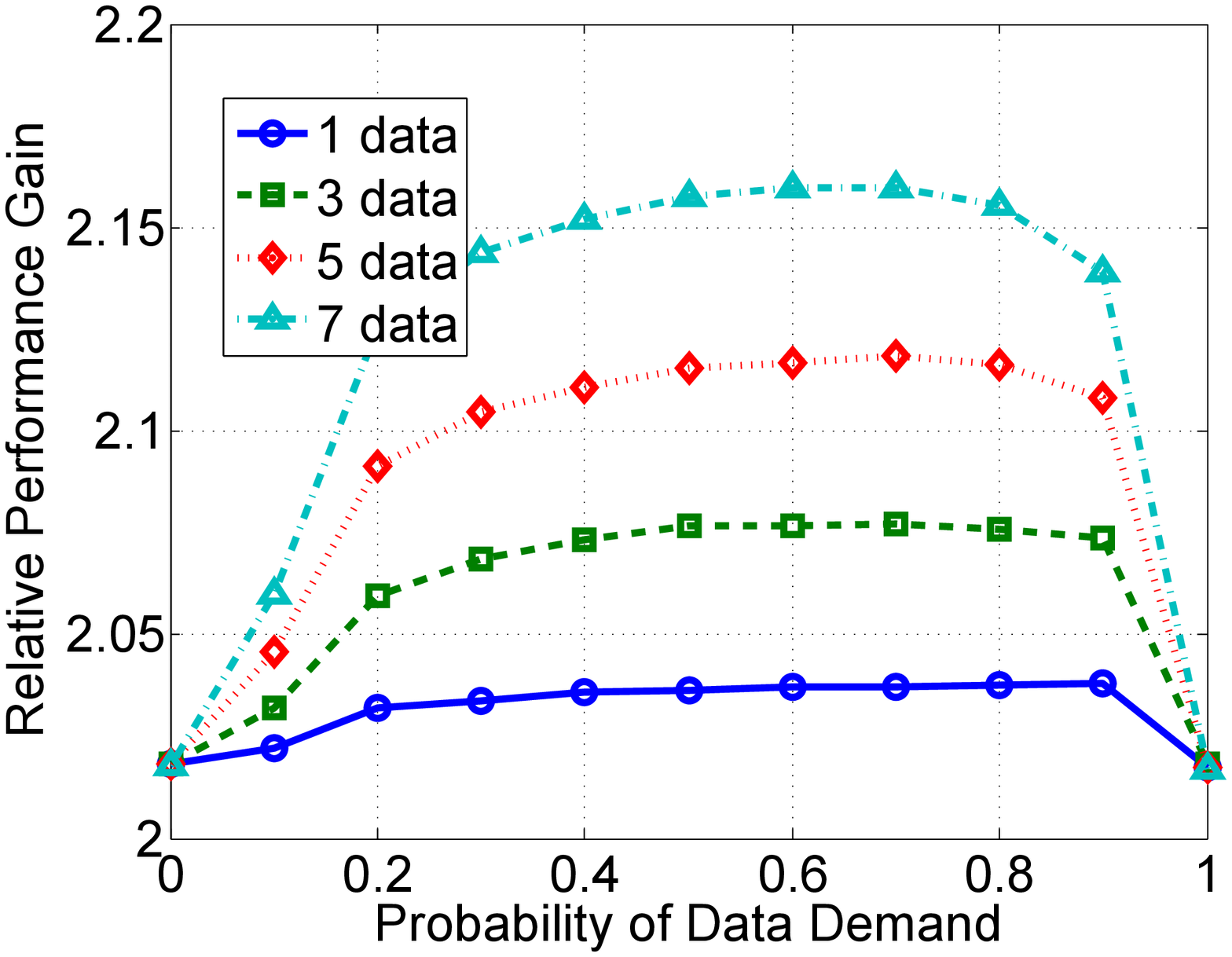}
  \caption{Impact of multiple data items on the relative performance gain}\label{Fig:MultipleData}
  \vspace{-4mm}
  \end{minipage}%
  \end{figure*}

\subsection{A Special Case of Two Tasks and Two Users}
The previous results are derived for large values of $I$ and $J$. Next, we will consider finite values of $I$ and $J$. As a special case, we will consider two tasks and two users, and derive some additional insights regarding data reuse. 

We assume two tasks and two users, and the task values and sensing costs follow i.i.d. uniform distributions on $[0,1]$. We will derive the \emph{explicit} expression for the gain $\gamma$.

We first consider the case without data reuse. The joint distribution of $g_{1,2:2}(c_1,c_2)$ is
$$g_{1,2:2}(c_1,c_2)=2g(c_1)g(c_2)=2,0\leq c_1<c_2\leq 1.$$
The joint distribution of $f_{1,2:2}(v_1,v_2)$ is
$$f_{1,2:2}(v_1,v_2)=2f(v_1)f(v_2)=2,1\geq v_1>v_2\geq 0.$$
Hence, the social welfare without data reuse can be computed by (\ref{eq:SOnoreuse}) as
$$
\begin{aligned}
& SW_{n}
 =4\int_{0}^{1}\!\!\int_{0}^{v_{1:2}}\!\!\!\!\int_{0}^{v_{1:2}}\!\!\!\!\int_{c_{1:2}}^{1}(v_{1:2}- c_{1:2})\d c_{2:2}\d c_{1:2}\d v_{2:2}\d v_{1:2}\\
& 
 +4\int_{0}^{1}\!\!\int_{0}^{v_{1:2}}\!\!\!\!\int_{0}^{v_{2:2}}\!\!\!\!\int_{c_{1:2}}^{v_{2:2}}(v_{2:2}\!-\! c_{2:2})\d c_{2:2}\d c_{1:2}\d v_{2:2}\d v_{1:2}=\frac{2}{5}.
\end{aligned}
$$

Now we consider the case with data reuse. We have 
$$f_{v_1+v_2}(v)=\min\{1,v\}-\max\{0,v-1\},0\leq v\leq 2,$$
and
$$f_{\min\{c_1,c_2\}}(c)=2(1-c),0\leq c\leq 1.$$  Hence, the social welfare with data reuse can be computed explicitly by (\ref{eq:SOreuse}) as
$$
\begin{aligned}
SW_{r} 
&=\int_{0}^{1}\int_{0}^{v}(v-c)\cdot v\cdot 2(1-c)\mathrm{d}c\mathrm{d}v\\
&~~+\int_{1}^{2}\int_{0}^{1}(v-c)\cdot (2-v)\cdot 2(1-c)\mathrm{d}c\mathrm{d}v=\frac{41}{60}.
\end{aligned}
$$

Hence, the relative performance gain due to data reuse is
$\gamma=SW_{r}/SW_{n}=\frac{41}{60}/\frac{2}{5}\approx 1.7.$

For the general case of a finite number of tasks and a finite number of users, we will use the \emph{Monte Carlo} method \cite{MonteCarlo} to compute $\gamma$ numerically. 
Figs. \ref{Fig:MonteCarlo} and \ref{Fig:MonteCarlo2} show the impact of the task number and the user number on the relative performance gain, respectively. We can see when the number of tasks equals the number of users (both are larger than 10), the relative performance gain is 2, which means that the social welfare with data reuse is doubled of that without data reuse. Furthermore, the relative performance gain is decreasing with the number of users, and increasing with the number of tasks.  Increasing the number of users has little impact on the social welfare with data reuse $SW_r$, since the user with the minimum sensing cost completes all tasks; while it can increase the social welfare without data reuse $SW_n$ due to the increasing user competition. Increasing the number of tasks can increase $SW_r$ due to data reuse; while it can decrease $SW_n$ due to the increasing task competition.

\subsection{Analysis for Multiple Data Items}
So far, we have considered the simplified scenario with one data item. The analysis for the scenario with multiple data items is quite challenging, due to the complicated coupling of tasks' data requirements and users' sensing capability across different data items. To show the key insights, we numerically study the impact of the number of data items on the performance gain with data reuse.

In the numerical studies, we fix both the number of tasks $J$ and the number users $I$ as $50$. The number of data items increases from 1 to 7. To show the impact of the number of data items on the relative performance gain $\gamma $, we assume that each task requires each data item with a fixed probability, and each user can senses each data item with the same probability. This captures the average data supply (users' sensing capabilities) and data demand (tasks' data requirements) for each data item.

Fig. \ref{Fig:MultipleData} shows the impacts of the number of data items and the data demand probability on the relative performance gain. We have two observations. First, we can see that the relative performance gain increases with the number of data items. On one hand, increasing the number of data items will decrease the social welfare with data reuse and that without data reuse,  due to the decreased task value per data and the increased sensing cost per data. On the other hand, allowing data reuse across tasks weakens the above effect, so that the reduction of the social welfare with data reuse is less that without data reuse. Hence, the relative gain increases with the number of data items. Second, the relative gain first increases and then decreases with the data demand probability. On one hand, the social welfare without data reuse first decreases and then increases with the data demand probability, due to the different impacts of task competition and user competition. When the probability is small, each task's data requirement is small and can be easily completed, leading to a larger social welfare. When the probability is large, each user's sensing capability is large, and the user competition leads to a larger social welfare.\footnote{When the probability is 0, all tasks (no data needs) can be completed without decreasing the social welfare. When the probability is 1, each task requires all data, and each user can sense all data. The above two cases are both equivalent to the case with one data item, and hence the gain is approximately 2, as is shown is Fig.  \ref{Fig:MonteCarlo}.} On the other hand, the social welfare with data reuse is approximately concave increasing with the data demand probability, due to the increasing reuse of data items. Hence, a larger relative reuse gain can be achieved when the data demand probability is medium.

However, theoretically understanding the benefit of data reuse is only the first step towards realizing the benefit of data reuse. In practice, tasks owners and users are selfish, and maybe unwilling to report their private information about task values and sensing cost. Hence we need to design an incentive mechanism to induce task owners and users to truthfully report their private information, while satisfying other properties such as achieving the maximum social welfare and computational efficiency.

\section{Auction-based Incentive Mechanisms}\label{sec:vcg}
In this section, we study the problem of achieving the above performance gain in the practical scenario with limited computational capability and incomplete information. We propose a two-sided auction-based incentive mechanism framework for solving Problem P1. First, we propose a two-sided VCG auction mechanism (as the benchmark) for solving Problem P1 exactly, which is feasible, socially optimal, but computationally difficult. Then we further propose a feasible, close-to-optimal, and low-complexity randomized auction mechanism for solving Problem P1 approximately in polynomial time. We aim to design an incentive mechanism satisfying the following five desirable properties:
\begin{itemize}
\item \textbf{Incentive Compatibility (IC, Truthfulness):} Reporting the true task value (and the true sensing cost, respectively) is the dominant strategy for each task owner (and each user, respectively), no matter what others report.
\item \textbf{Individual Rationality (IR):} Each participating task owner and user will have a non-negative utility by reporting the true task value and sensing cost, respectively.
\item \textbf{Feasibility and Economic Efficiency:} The outcome of the mechanism can be implemented in practice (i.e., through an integer allocation) and maximizes the social welfare.
\item \textbf{Computational Efficiency:} The outcome of the mechanism can be computed in polynomial time.
\item \textbf{Budget Balance:} The total payment obtained from the selected task owners should be no less than the total payment paid to the scheduled users.
\end{itemize}

\subsection{Two-sided Auction Framework}\label{subsec:auctionframework}

To solve Problem P1 with two-sided private information, we propose a two-sided auction-based incentive mechanism, where the platform acts as an \emph{auctioneer} purchasing data from mobile users (\emph{bidders} on one side) and selling data to task owners (\emph{bidders} on the other side).
In this auction framework,
the platform first announces an \emph{allocation rule} (for task selection and user scheduling) and a \emph{payment rule} (for payments to the scheduled users \emph{and} prices charged to the selected task owners).
Then, each task owner submits a bid (indicating her task value) and each user submits a bid  (indicating her sensing cost) to the platform, which can be different from the true task value and the true user sensing cost, respectively.
Finally, the platform computes the allocations and payments,
based on the reported bids of all task owners and users, together with other public information  (e.g., tasks' data requirements and users' sensing capabilities).
In this work, we are interested in designing the \emph{truthful} auction, where task owners and users submit their private information truthfully.

Next, we provide the key notations. Let $u_j$ denote the reported value (bid) of task~$j$. Let $\b_i = (b_i(\S), \ \S \subseteq\S_i)$ denote the reported sensing cost vector (bid) of user $i$, where $b_i(\S)$ denotes the user reported sensing cost for a data set $\S \subseteq \S_i$.
Let $\bv\eq (u_j,j\in\J)$ denote the bids of all tasks and $\b  \eq (\b _i, i\in\I) $ denote the bids of all users.
If an auction is truthful, we will have $\b  = \c $ and $\bv=\v$ at the equilibrium.
With a little abuse of notation, we denote  $\{\x (\cdot), \z (\cdot)\} $ as the allocation rule, where
$\x (\cdot) \eq (\x _i(\cdot),i\in\I)$ is the user scheduling vector and
$\z (\cdot) \eq (z_j(\cdot),j\in\J)$ is the task selection vector. We further denote $\{\p (\cdot),\pv(\cdot)\}$ as the payment rule, where $\p(\cdot)\eq (p_i(\cdot),i\in\I)$ is the user payment vector and $\pv(\cdot)\eq (q_j(\cdot),j\in\J)$ is the task charge vector.
Note that $\x (\cdot)$, $\z (\cdot)$, $\p (\cdot)$, and $\pv (\cdot)$
can also be written as $\x (\b,\bv)$, $\z (\b,\bv)$, $\p (\b,\bv)$, and $\pv (\b,\bv)$, as they  are all functions of the user bid vector $\b$ and the task bid vector $\bv$.
For convenience, we write such an auction mechanism as $\mech \eq \{ \x (\cdot), \z (\cdot); \p (\cdot),\pv (\cdot) \}$ or  $\mech \eq \{ \x (\b,\bv), \z (\b,\bv); \p (\b,\bv), \pv (\b,\bv)\}$. ~~~~~~~~~~~~~~~~~~~~~~~~~~~~~

\subsection{Two-sided VCG Auction (Benchmark)}\label{subsec:VCG}

We first propose a two-sided VCG auction, which is a nontrivial extension of the classic VCG auction~\cite{vcg}, due to the two-sided information asymmetry. In our two-sided VCG  auction, the allocation rule aims to maximize the social welfare \emph{defined on the reported sensing costs and task values}, and the payment rule aims to pay each scheduled user the social benefit that she generates \emph{and} to charge each selected task owner the social damage that she imposes. Formally, 

\begin{mechanism}[Two-sided VCG Auction Mechanism -- $\mech^o$]\label{mech:1}
~
\begin{itemize}
\item
Allocation Rule $\{\x   (\b,\bv),\z   (\b,\bv)\}$:
$$
\x  (\b,\bv) = \x ^o(\b,\bv ) \mbox{~~and~~}
\z   (\b,\bv)= \z ^o(\b,\bv),
$$
where $\{\x ^o(\b,\bv ), \z ^o(\b,\bv )\}$ is the optimal solution to Problem P1, by replacing $\c $ with the reported cost $\b $ and $\v$ with the reported value $\bv$ in Problem P1;
\item
Payment Rule $\{\p   (\b,\bv),\pv   (\b,\bv)\}$:
\begin{gather}
\p   (\b,\bv) = \p^o  (\b,\bv) \eq (p_i^o (\b,\bv))_{i\in\I},\notag\\
\pv   (\b,\bv) = \pv^o  (\b,\bv) \eq (q_j^o (\b,\bv))_{j\in\J},\notag
\end{gather}
where
\begin{gather}
\textstyle p_i^o  (\b,\bv) \!\!\eq\!\!
 \sum\limits_{j\in \J}  u_j z_j^o (\b,\bv )
\!\!- \!\!\!\!\!\sum\limits_{n \in \I \setminus \{i\}} \!\sum\limits_{\S \subseteq \S_n} \!\! b_n (\S) x_n^o (\S)  \!\!-\!\! W^o_{-i} ,\notag\\
\textstyle q_j^o  (\b,\bv) \!\!\eq\!\!  W^o_{-j} \!\!-\!\!\!\!\!
 \sum\limits_{j\in \J \setminus \{j\}} \!\!\! u_j z_j^o (\b,\bv )
\!\!+\!\! \sum\limits_{n \in \I } \sum\limits_{\S \subseteq \S_n} \!\! b_n (\S) x_n^o (\S) ,\notag
\end{gather}
where $W^o_{-i}$ is the maximum social welfare (defined on bids $\b,\bv $) excluding user $i$'s bid; $W^o_{-j}$ is the maximum social welfare (defined on bids $\b,\bv $) excluding task $j$'s bid.\footnote{Specifically, $W^o_{-i}  $ is the maximizer of Problem P1, by replacing $\c$ with $\b$ and $\v$ with $\bv$, and excluding user $i$'s bid $\b_i$, before solving Problem P1. Similarly, $W^o_{-j}$ is obtained by excluding task $j$'s  bid $u_j$.}
\end{itemize}
\end{mechanism}

In Mechanism \ref{mech:1}, task owner $j\in\J$ chooses the bid $u_j^{o}$ such that $u_j^{o}=\arg\max_{u_j} ( v_j-q_j^o  (\b,\bv))$; user $i\in\I$ chooses the bid $\b_i^{o}$ such that $\b_i^{o}=\arg\max_{\b_i} (p_i^o  (\b,\bv)- \sum_{\S \subseteq \S_i }c_i(\S))$. The bid $(\b^{o},\bv^{o})$ is a Nash equilibrium, if each   user and each task owner have no incentives to unilaterally change her bid, respectively.
For convenience, we write Mechanism \ref{mech:1} as $ \mech^o =  \{\x^o (\cdot), \z^o(\cdot); \p^o(\cdot),\pv^o(\cdot)\} $ or $ \mech^o =\{\x^o (\b,\bv), \z^o(\b,\bv); \p^o(\b,\bv),\pv^o(\b,\bv)\} $. 
By extending the analysis of the standard VCG auction \cite{vcg} to our \emph{two-sided scenario}, we can show that truthful reporting is a dominant strategy for both users and task owners, i.e., $\b^{o}  = \c $ and $\bv^{o}=\v$ constitute the unique Nash equilibrium.
This further implies that Mechanism~\ref{mech:1} is efficient, as its allocation maximizes the social welfare defined in~\eqref{eq:Sxz}.~~~~~~~~~

\begin{proposition}[Truthfulness and Efficiency]\label{prop:VCG}
Mechanism~\ref{mech:1} is individually rational, incentive compatible (truthful), and   maximizes the social welfare (efficient).
\end{proposition}

Although Mechanism~\ref{mech:1} exhibits several desirable properties, computing the two-sided VCG auction outcome needs to solve the NP-hard Problem P1, which is computationally intractable.
To this end, we will propose a low-complexity auction mechanisms next.

 \subsection{Two-sided Randomized Auction}
\label{subsec:auction}

Inspired by the randomized auction in \cite{ra,RAuction}, we now propose a low-complexity two-sided randomized auction mechanism, which operates in polynomial time. Due to the two-sided structure of mutual information asymmetry, our auction is different from the traditional single-sided auctions \cite{ra,RAuction}.  

In the following, we start from the linear programming relaxation of Problem P1, obtain an associated linear programming Problem P2 \emph{in the fractional domain}, from which we further derive the fractional VCG auction (which may not be implementable in practice).
Then, through proper decompositions, we transform the fractional VCG auction to a two-sided randomized auction (which is implementable).

\subsubsection{Linear Programming Relaxation}

We first relax the joint task selection and user scheduling Problem P1 to the fractional domain (i.e., relax every binary variable in $\{0, 1\}$ to the domain $[0,1]$), and denote the associated linear programming problem as Problem P2.
Note that Problem P2 can be solved to its optimality  in polynomial time, as it is a standard linear programming problem\cite{lp}.
We refer to the optimal solution of Problem P2 as the \emph{fractional optimal solution}, denoted by
$\{\x ^\ast, \boldsymbol{y}^\ast, \z ^\ast\}$ or $\{\x ^\ast (\boldsymbol{c},\v), \boldsymbol{y}^\ast(\boldsymbol{c},\v), \z ^\ast(\boldsymbol{c},\v)\}$.
It is notable  that the maximum objective value of Problem P2 provides an upper-bound for the optimal objective function value of Problem P1, and the gap is usually called the \emph{integrality gap} \cite{lp}.
Intuitively, a fractional solution can be viewed as \emph{the fraction of the time} that users are scheduled or tasks are selected.

Next, we present the fractional VCG auction $\mech^\ast$, where the allocation rule aims to maximize the social welfare (based on user bids $\b $ and task bids $\bv $)  \emph{in the fractional domain},
and the payment rule aims to pay each scheduled user her social benefit and charge each selected task her social damage. The detailed mechanism is similar to $\mech^o$, except that we replace the integer solution $\{\x ^o(\b,\bv ),\z ^o(\b,\bv )\}$ by the fractional optimal solution $\{\x ^\ast(\b,\bv ),\z ^\ast(\b,\bv )\}$, and solving Problem P2 rather than P1 when deciding the payments. We formally show the mechanism as follows.

\begin{mechanism}[Fractional VCG Auction Mechanism -- $\mech^\ast$]\label{mech:2}
 ~
 \begin{itemize}
 \item
 Allocation Rule $\{\x  (\b,\bv ),\z   (\b,\bv )\}$:
 $$
 \x (\b,\bv ) = \x ^\ast(\b,\bv ) \mbox{~~and~~}
 \z  (\b ,\bv)= \z ^\ast(\b,\bv ),
 $$
 where $\{\x ^\ast(\b,\bv ), \z ^\ast(\b,\bv )\}$ is the optimal solution to Problem P2, by replacing $\boldsymbol{c}$ with the reported cost $\b $ and $\v$ with the reported value $\bv$ in Problem P2;
 \item
 Payment Rule $\{\p   (\b,\bv),\pv   (\b,\bv)\}$:
\begin{gather}
 \p   (\b,\bv) = \p^\ast  (\b,\bv) \eq (p_i^\ast (\b,\bv))_{i\in\I},\notag\\
\pv   (\b,\bv) = \pv^\ast  (\b,\bv) \eq (q_j^\ast (\b,\bv))_{j\in\J},\notag
\end{gather}
 where
\begin{gather}
\textstyle p_i^\ast (\b,\bv) \!\!\eq\!\!
 \sum\limits_{j\in \J}  u_j z_j^\ast (\b,\bv )
\!\!- \!\!\!\!\!\sum\limits_{n \in \I \setminus \{i\}} \!\sum\limits_{\S \subseteq \S_n} \!\! b_n (\S) x_n^\ast (\S)  \!\!-\!\! W^\ast_{-i} ,\notag\\
\textstyle q_j^\ast  (\b,\bv) \!\!\eq\!\!  W^\ast_{-j} \!\!-\!\!\!\!\!
 \sum\limits_{j\in \J \setminus \{j\}} \!\!\! u_j z_j^\ast (\b,\bv )
\!\!+\!\! \sum\limits_{n \in \I } \sum\limits_{\S \subseteq \S_n} \!\! b_n (\S) x_n^\ast (\S) ,\notag
\end{gather}
where $W^\ast_{-i}  $ is the maximum social welfare (defined on bids $\b,\bv $) excluding user $i$'s bid in the fractional domain, and $W^o_{-j}$ is the maximum social welfare (defined on bids $\b,\bv $) excluding task $j$'s bid in the fractional domain.
 \end{itemize}
 \end{mechanism}

We summarize the properties of Mechanism~\ref{mech:2} as follows.

\begin{proposition}[Truthfulness and Efficiency]\label{prop:fractionalVCG}
Mechanism~\ref{mech:2} is individually rational, incentive compatible (truthful), and maximizes the social welfare (efficient) in the fractional domain.
\end{proposition}

Note that the optimal solution to Problem P2 (i.e., the outcome of Mechanism~\ref{mech:2}) may \emph{not} be feasible to   Problem P1.
This implies that \emph{Mechanism~\ref{mech:2} may not be implementable}.
To see this, consider an example with 3 data items $\D=\{1,2,3\}$, 3 users $\I = \{1,2,3\}$ with sensing capabilities $\S_1=\{1,2\}$, $\S_2=\{1,3\}$, and $\S_3=\{2,3\}$, and 1 task requiring all of 3 data items.
Suppose that the user's sensing capability is single-minded, i.e., each user $i$ either senses all the data items in $\S_i$ or does not sense  any data item.
Then, the fractional optimal solution is to schedule each user half of the time, i.e., $x_1^*(\S_1) = x_2^*(\S_2) =  x_3^*(\S_3) = 0.5$, and to complete the task all the time, i.e., $z_1^* = 1$. However, such a fractional solution cannot be implemented in practice, since each user should be either selected or not selected.
In the following, we will transform Mechanism~\ref{mech:2}, i.e., the fractional VCG auction~$\mech^\ast$, to a randomized auction, which always generates a feasible solution to Problem P1 randomly according to certain probability, hence is implementable.

\subsubsection{Randomized Mechanism Definition}
We first provide the definition of a randomized mechanism and the associated concept of truthfulness in expectation~\cite{ra}.

Recall that a two-sided \emph{deterministic} mechanism $\mech = \{\x (\cdot), \z (\cdot); \p(\cdot),\pv(\cdot)\}$  consists of a deterministic  allocation rule $\{\x (\cdot), \z (\cdot)\}$ and a payment rule $\{\p (\cdot),\pv(\cdot)\}$, and returns a deterministic  outcome $\{\x (\b,\bv ), \z (\b,\bv ); \p (\b,\bv ),\pv(\b,\bv)\}$ given any bids $\b $ and $\bv$.
Note that both Mechanism \ref{mech:1} and Mechanism \ref{mech:2} introduced before are deterministic mechanisms.

A mechanism $\wmech \eq  \{\wx (\cdot), \wz (\cdot); \wp (\cdot),\wpv(\cdot)\}$ can also be randomized, 
in which the allocation and payment determinations involve randomizations. 
In other words, given any bids $\b $~and~$\bv$, the outcomes $\widetilde{x}_i(\b,\bv )$, $\widetilde{z}_j(\b,\bv  )$, $\widetilde{p}_i(\b,\bv)$ and $\widetilde{q}_j(\b,\bv)$ are all random variables. As the result, each task owner's utility (i.e., value minus charge) and each user's utility (i.e., payment minus sensing cost) are also random variables.
Intuitively, such a randomized mechanism can be viewed as a set of randomizations over the deterministic mechanism.
For randomized mechanisms, the concept of truthfulness is defined in the expected sense.
That is, if a randomized mechanism $\wmech$ is truthful in expectation, then the \emph{expected} utilities of each user and each task owner are maximized when reporting the true sensing cost and value, respectively. 

\subsubsection{Randomized Mechanism Design Criterion}
We now provide our design criterion of a \emph{truthful} randomized mechanism.
The key idea is to find a randomized mechanism that generates the equivalent outcome of a truthful deterministic mechanism.

We first introduce an $(\alpha,\beta)$-scaled fractional mechanism  for the deterministic mechanism $ \mech = \{\x (\cdot), \z(\cdot); \p(\cdot),\pv(\cdot) \}$, inspired by the $\alpha$-scaled fractional mechanism defined in \cite{ra,RAuction}. Comparing with the one-sided mechanisms in \cite{ra,RAuction}, our mechanism considers the scaling of both sides.

\begin{definition} [Scaled Fractional Mechanism]
An $(\alpha,\beta)$-scaled fractional mechanism of $\mech =\{\x (\cdot), \z(\cdot);  \p(\cdot),\pv(\cdot)\}$, denoted by
 $\mech_{(\alpha,\beta)} = \{\x_\alpha(\cdot), \z_\beta(\cdot); \p_\alpha(\cdot),\pv_\beta(\cdot)\}$, is defined as:
\begin{gather}
 \textstyle
\x_\alpha(\cdot)  = {\alpha} \cdot  \x (\cdot),\ \
\p_\alpha (\cdot) = {\alpha}  \cdot \p(\cdot),\\
 \z_\beta(\cdot)  = {\beta} \cdot  \z (\cdot), \ \
\pv_\beta (\cdot) = {\beta}  \cdot \pv(\cdot),
\end{gather}
where $ {\alpha},{\beta} > 0$ are the scaling factors such that every element of ${\alpha} \cdot  \x (\cdot)$ belongs to $[0,1]$ and every element of ${\beta} \cdot  \z (\cdot)$ belongs to $[0,1]$, respectively.
\end{definition}

Intuitively, in an $ ({\alpha},\beta)$-scaled fractional mechanism, the incurred cost and payment of each user are scaled with~$\alpha$, and the achieved value and charge of each task owner are scaled with $\beta$, compared with those in the original mechanism~$\mech$.
This implies that both the users' and the task owners' optimal bidding strategies in these two mechanisms are equivalent, which leads to the equivalence of the truthfulness property of both mechanisms. 

\begin{proposition}\label{prop:xx1}
If a mechanism $\mech $ is truthful, then its $({\alpha},\beta)$-scaled fractional mechanism
$\mech_{(\alpha,\beta)} $ is also truthful.
\end{proposition}

Based on the Proposition \ref{prop:xx1}, we propose the following two-sided randomized mechanism design criterion. That is, design a  two-sided randomized mechanism $\wmech$ that provides the equivalent outcome (in terms of task selection, user scheduling, and payment) as an $({\alpha},\beta)$-scaled fractional mechanism $\mech^*_{({\alpha},\beta)} $ of the fractional VCG auction~$\mech^*$ in Mechanism~\ref{mech:2}.

As the fractional VCG auction mechanism $\mech^*$ in Mechanism \ref{mech:2} is truthful, we can obtain the truthfulness of its $({\alpha},\beta)$-scaled fractional mechanism by Proposition~\ref{prop:xx1}.
Moreover, as the randomized mechanism $\wmech $ generates the same task selection, user scheduling, and payment as $\mech^*_{({\alpha},\beta)} $,
we can further obtain the truthfulness (in expectation) of $ \wmech$.

\subsubsection{Two-sided Randomized Mechanism Design}
Now we provide the details about our two-sided randomized mechanism design.

For convenience, we express a randomized mechanism $
\wmech =  \{\wx(\cdot), \wz(\cdot); \wp (\cdot),\wpv (\cdot)\} $ as a set of allocation probabilities $\boldsymbol{\lambda} = (\lambda^l)_{l\in \A}$ and a set of payment rules $\{\p^l(\cdot), \pv^l(\cdot) \}_{l\in\A}$ under all possible allocations,
 where $\A$ is the set of all feasible \emph{integer} allocations (regarding $ \x $ and $ \z $) and $\lambda^l \geq 0$ is the probability of picking up a particular allocation $\{\x^l, \z^l\}$ and the corresponding payment $ \{\p^l, {\pv}^l\}$.
Then, designing a randomized mechanism $\wmech $ is equivalent to finding a set of allocation  probabilities $\boldsymbol{\lambda} = (\lambda^l)_{l\in \A}$ and a set of payment rules $\{ \p^l (\cdot),{\pv}^l(\cdot)\}_{l\in\A}$.

Next, we propose the two-sided randomized auction~$\wmech^\dag$, which aims to maximize the
two-sided scaled social welfare subject to the exact decomposition of the fractional optimal solution into the weighted sum of integer solutions.
Due to the two-sided social welfare maximization, $\wmech^\dag$ nontrivially  extends those with one-sided utility maximization or cost minimization in \cite{ra,RAuction}.

\setcounter{mechanism}{2}
\begin{mechanism}[Randomized Auction Mechanism -- $\wmech^\dag$]\label{mech:3}~

Starting from the fractional VCG auction $\mech^* = \{\x^\ast(\b,\bv) ,$ $\z^\ast(\b,\bv); \p^\ast(\b,\bv),{\pv}^\ast(\b,\bv)\}$ in Mechanism \ref{mech:2}, we define the randomized auction mechanism $\wmech^\dag$ as:
\begin{itemize}
\item
Allocation Rule  $\boldsymbol{\widetilde{\lambda}} = (\lambda^l)_{l\in \A}$:
\begin{align}
\textstyle \boldsymbol{\widetilde{\lambda}}  =  &  \arg \max_{\boldsymbol{\lambda}, 0<\alpha,\beta\leq 1} ~~ \beta\cdot V^\ast-\alpha\cdot C^\ast
\\
& \textstyle \mbox{s.t.}\textstyle  \sum_{l \in \A}
\lambda^l \cdot
\x_i^l = \alpha  \cdot\x_i^\ast (\b,\bv) , \quad \forall i\in \I,\label{eq:mechconstraint1}\\
& ~~~~\textstyle \sum_{l \in \A}
\lambda^l \cdot
z_j^l  = \beta  \cdot  z_j^\ast (\b,\bv), \quad \forall j\in \J,\label{eq:mechconstraint2}
\end{align}
where $V^\ast$ and $C^{\ast}$ are the optimal total task values and user costs w.r.t. $\z^\ast(\b,\bv)$ and $\x^\ast(\b,\bv)$, respectively.
\item
Payment Rule  $\{\p^l (\b,\bv),{\pv}^l(\b,\bv) \}_{l\in \A} $:
\begin{gather}
\textstyle
p_i^{l} (\b,\bv) = \alpha \cdot p_i^\ast (\b,\bv) \cdot \frac{C_i(\x_i^l)}{ \sum_{l' \in \A}
\lambda_{l'} \cdot
C_i(\x_i^{l'}) },\forall i\in\I,\notag\\
\textstyle
{q}_j^{l} (\b,\bv) = \beta \cdot q_j^\ast (\b,\bv) \cdot \frac{V_j(z_j^l)}{ \sum_{l' \in \A}
	\lambda_{l'} \cdot
	V_j(z_j^{l'}) },\forall j\in\J,\notag
\end{gather}
where $C_i(\x_i^l)$ is user $i$'s cost under the allocation $\x_i^l$, and $V_j(z_j^l)$ is task $j$'s value under the allocation $z_j^l$.
\end{itemize}
\end{mechanism}

We can see that in  Mechanism \ref{mech:3}, both the expected payment and sensing cost of each user \emph{and} the expected charge and value of each task are equivalent to those in the fractional VCG auction $\mech^*$ in Mechanism \ref{mech:2}, which implies that Mechanism \ref{mech:3} is truthful \emph{in expectation}. 

\begin{proposition}[Incentive Compatibility in Expectation]\label{prop:ratruthful}
Mechanism \ref{mech:3} is incentive compatible in expectation, in the sense that each user and task owner can maximize her expected utility when reporting the true sensing cost and value, respectively.
\end{proposition}

We can further check that under Mechanism \ref{mech:3}, each user and task owner can always achieve a non-negative utility \emph{under any possible realization of allocations}. This implies that
Mechanism \ref{mech:3} is individually rational \emph{in the strict sense}. Formally,

\begin{proposition}[Individual Rationality]
Mechanism \ref{mech:3} is individually rational in the strict sense, as each user and task owner can achieve a non-negative expected utility.
\end{proposition}

Furthermore, we can see that in Mechanism \ref{mech:3}, each user's sensing cost equals $\alpha^*$ times the sensing cost incurred in Mechanism \ref{mech:2}, while each task's value equals~$\beta^*$ times the value achieved in Mechanism \ref{mech:2} (where $\alpha^*$ and $\beta^*$ are the optimal solutions to the allocation problem in Mechanism~\ref{mech:3}).
Hence,   the efficiency of   Mechanism \ref{mech:3} is guaranteed in this sense.~~~~~

\begin{proposition}[Efficiency of Mechanism $\wmech^\dag$]
Mechanism \ref{mech:3} guarantees to achieve a $\beta^*$-fraction of the total task value in Mechanism \ref{mech:2} with an $\alpha^*$-fraction of the total sensing cost in Mechanism \ref{mech:2}.
\end{proposition}

So far we have proposed the randomized auction mechanism and proved several desirable economic properties. There are many possible ways to implement the randomized auction, depending on how we obtain the set of probability distribution for the allocation problem and the parameter $\alpha$ and $\beta$ for the payment in Mechanism \ref{mech:3}. Next, we will propose one easy-to-implement solution method.

\subsubsection{Implementation of the Randomized Auction}
As we have mentioned, one key step of designing Mechanism~\ref{mech:3} is the two exact decompositions of the scaled fractional solutions into the weighted sum of integer solutions in (\ref{eq:mechconstraint1}) and (\ref{eq:mechconstraint2}), to obtain the two scaling factors $\alpha$ and $\beta$. Next, we will show that it may sacrifice some social welfare in order to achieve the exact decompositions efficiently, which is a key difference of our approach here and the approach proposed in \cite{RAuction}. In \cite{RAuction}, the authors proposed a decomposition method to ensure system efficiency, but with a very complicated procedure that may not be practical.\footnote{For example, the ellipsoid method used in \cite{RAuction} is quite complicated, and incurs a high time complexity in practical systems.} Next, we exploit our two-sided problem structure to obtain a tailored easy-to-implement decomposition.\footnote{This is just one of the many possible solutions, which may differ in computational complexity and system efficiency loss.}

The key idea of the our solution approach is a two-step \underline{DE}composition-\underline{MO}dification (\textsf{DEMO}) procedure, which is shown as follows.
\begin{itemize}
\item Step 	I: \emph{Decomposition}. We start from the fractional optimal solution $(\x^*,\z^*)$. Given the fractional user scheduling solution  $\x^*$, we treat the fractional $\x^*$ as the probability of scheduling the corresponding user. More specifically, we propose the following approach to compute $\lambda_l$. First recall that a feasible integer user scheduling is $\x^l=\{x_i^l(\S),\forall\S\subseteq\S_i,\forall i\in\I\}$ and the fractional optimal user scheduling is $\x^*=\{x_i^*(\S),\forall\S\subseteq\S_i,\forall i\in\I\}$. Then we define the probability distribution $\lambda^l$ as
\begin{equation}\label{eq:decomposition}
\lambda^l=\prod_{i\in\I}\prod_{\S\subseteq\S_i}\phi(x_i^*(\S),x_i^l(\S)),\forall l\in\A,
\end{equation}
where $\phi(x_i^*(\S),x_i^l(\S))=x_i^*(\S)$ if $x_i^l(\S)=1$, and $\phi(x_i^*(\S),x_i^l(\S))=1-x_i^*(\S)$ if $x_i^l(\S)=0$. The function $\phi(x_i^*(\S),x_i^l(\S))$ characterizes the probability of scheduling user $i$ with the set $\S$, and the probability is given by the corresponding fractional solution.   
\item Step 	II: \emph{Modification}. Each integer solution $\x^l$ corresponds to a maximum set of task selection $\z^l$. Given the probability distribution $\lambda^l,\forall l\in\A$, we compute 
\begin{equation}
\beta_j=\frac{\left(\sum_{l \in \A}\lambda^l \cdot z_j^l\right)}{z_j^\ast}, \forall j\in\J.
\end{equation}
In order to ensure that all $z_j^\ast$ are scaled by the same factor $\beta$, we choose the smallest value of $(\beta_j,j\in\J)$ as the target $\beta$. Then we modify $(z_j^l,j\in\J)$ such that $\beta_j=\beta,\forall j\in\J$. The detailed modification approach is as follows. First notice that each user scheduling $\x^l$ corresponds to the maximum set of the task selection $\z^l=(z_j^l,j\in\J)$. Then for a particular $z_{\jmath}^l=1$, we can modify $z_{\jmath}^l=1$ as $z_{\jmath}^l=0$ without violating the feasibility of the task selection. We iteratively conduct the above modification procedure until we have $\beta_j=\beta,\forall j\in\J$.
\end{itemize}
We can show the probability distribution $\lambda^l$ in (\ref{eq:decomposition}) satisfies
$\x^*=\sum_{l\in\A}\lambda^l\boldsymbol{x}^{l}$
with $\alpha=1$. Then, we can choose $\beta$ according to the modification procedure in Step II such that
$\sum_{l \in \A}\lambda^l \cdot z_j^l  = \beta  \cdot  z_j^\ast,\forall j\in\J.$

Through the above proposed \textsf{DEMO} scheme, we derive the target $\lambda^l$, $\alpha^{\ast}$, and $\beta^{\ast}$. That is, we obtain the exact decompositions of the scaled fractional solutions into the weighted sum of integer solutions as in (\ref{eq:mechconstraint1}) and (\ref{eq:mechconstraint2}). According to Propositions \ref{prop:xx1} and \ref{prop:ratruthful}, we have ensured the truthfulness of the mechanism at the cost of a reduced system efficiency.

\begin{proposition}[Truthfulness and Efficiency Bound]
The \textsf{DEMO} procedure implements Mechanism~\ref{mech:3}, and guarantees to achieve the same total sensing cost (i.e., $\alpha^{\ast}=1$) in Mechanism \ref{mech:2} with a $\beta^{\ast}$-fraction of the total task value in Mechanism~\ref{mech:2}, where 
\begin{equation}
\textstyle \beta^{\ast}=\min\limits_{ j\in\J}\left(\frac{\sum_{l \in \A}\lambda^l \cdot z_j^l}{z_j^\ast}\right).
\end{equation}
\end{proposition}

We have proved several desirable properties of our designed auction mechanisms. Due to the two-sided structure of the auction mechanisms, the platform may lose money if the total payment obtained from task owners is less than the total payment paid to users. In other words, Mechanism \ref{mech:3} may not be budget-balanced, which can be a practical concern. In fact, it is a well known result in the literature that truthful efficient mechanisms may not be budget-balanced \cite{EfficiencyBudget,EfficiencyBudget2}.
Next, we further focus on the budget-balanced auction mechanism design.

\section{Budget-Balanced Randomized Auction}\label{sec:extension}
In this section, we focus on the budget-balance property of Mechanism \ref{mech:3}, i.e.,  the expected total payment paid to users should be no large than the expected total payment obtained from task owners. This means that the MCS platform will not lose money, which is quite important for the realistic operation of the MCS platform. Since the expected payments in Mechanism \ref{mech:3} for all users and task owners are scaled from Mechanism \ref{mech:2} by the same factors $\alpha$ and $\beta$, we first focus on the budget balance of Mechanism \ref{mech:2}, and then extend the results to Mechanism \ref{mech:3}.

\subsection{Budget Balance}
In our model, we say that a mechanism is budget-balanced, if the MCS platform can achieve a non-negative profit, where the MCS platform's profit is defined as the difference between the total payments obtained from task owners and the total payment paid to users. Based on our above discussion, the budget balance of the two-sided auction cannot by guaranteed in general. In particular, with the increase of the task similarity, the positive network effect among tasks also increases, and the total payments from task owners to the platform become smaller and smaller (even zero). In such cases, the budget-balance property is not satisfied.

We use an illustrative example to show the budget imbalance. Suppose we have four tasks (tasks 1-4) with task values 0.5, 0.6, 0.7, and 0.8, respectively. Each task only requires the same one data item. Two users (users 1-2) can sense the data item, with sensing costs 0.1 and 0.2, respectively. The social welfare maximizer would require user 1 to sense the data item with the cost 0.1, and allow all four tasks to reuse the data item. Hence, the maximum social welfare is 0.5+0.6+0.7+0.8-0.1=2.5. Now we consider Mechanism \ref{mech:2}. The VCG auction would schedule user 1 to sense the data item with a payment 0.2, and all tasks would be selected. The payment of task 1 would be $(0.5+0.6+0.7+0.8-0.1-0.2+0.2-0.5)-(0.6+0.7+0.8-0.1-0.2+0.2) =0,$ where the first term is the total social welfare except task 1 (when task 1 is considered in the auction), and the second term is the total social welfare when excluding task 1 from the auction.
Similarly, we can show that the payments of tasks 2-4 are all zero. Hence, the total payment from task owners to the platform is 0, while the total payment from the platform to the sensing user is 0.2. This shows that Mechanism \ref{mech:2} may not be budget-balanced. Due to the scaled payments from Mechanism \ref{mech:2}, Mechanism \ref{mech:3} is not guaranteed to be budget-balanced either.

\subsection{Reserve Price based Randomized Auction}\label{reservepriceauction}
In the following, we will first focus on making Mechanism~\ref{mech:2}  budget-balanced, and then extend the results to the budget-balanced randomized auction design by scaling the payments in Mechanism~\ref{mech:2} according to the scaling rule proposed in Mechanism \ref{mech:3}.

To this end, we introduce a reserve price for each data item in the proposed Mechanism \ref{mech:2}, which denotes the minimal payment that a task owner has to pay for each data item. Let $\sigma_k\geq0$ denote the reserve price for each data item $k\in\K$.
Then, for each task owner $j \in\J$, the minimum payment (if task $j$ is completed) is
\begin{equation}\label{eq:reservepricedata}
\textstyle \underline{q}_j^\sigma=\sum_{k\in\K_j}\sigma_k.
\end{equation}
Given the above minimum payments $(\underline{q}_j^\sigma, j\in\J)$ due to the reserve price $\sigma_k$, to ensure truthfulness, we propose the following bids reduction and payment rule for task owners. 
\begin{definition}[Bids Reduction and Payment Rule]
Given users' bids $\b$, task owners' bids $\bv$, and the minimum payments $(\underline{q}_j^\sigma, j\in\J)$, the reduced bid of task owner $j$ is $u'_j=u_j-\underline{q}_j^\sigma.$ With the new reduced bids $\bv'=(u'_j,j\in\J)$, Mechanism~\ref{mech:2} leads to the allocation $(\x ^\sigma(\b,\bv'),\z ^\sigma(\b,\bv'))$ and the payment $(\p^\sigma(\b,\bv'),{\pv}^\sigma(\b,\bv'))$. Then the payment of task owner $j\in\J$ is
\begin{equation}\label{eq:reserveprice}
q_j^\sigma (\b,\bv)=q_j^\sigma (\b,\bv')+\underline{q}_j^\sigma.
\end{equation}
\end{definition}

The key idea of proposing the above bids reduction and payment rule is to reduce the mechanism design problem to a setting with no minimum payments. In particular, we first subtract the minimum payment of each task owner from her bid, run Mechanism \ref{mech:2}, and then add the minimum payment of each task owner to her resulting payment.

Next, we propose the two-sided randomized auction mechanism $\wmech^\sigma$ in Mechanism \ref{mech:reserveprice}, i.e.,  the two-sided randomized auction mechanism with the reserve price.
\begin{mechanism}[Randomized Auction Mechanism with the Reserve Price -- $\wmech^\sigma$]\label{mech:reserveprice}
 ~
 \begin{itemize}
 \item
Allocation Rule  $\boldsymbol{\widetilde{\lambda}} = (\lambda^l)_{l\in \A}$:
 \begin{equation*}
 \begin{aligned}
 \textstyle \boldsymbol{\widetilde{\lambda}}  =  &  \arg \max_{\boldsymbol{\lambda}, 0<\alpha,\beta\leq 1} ~~ \beta\cdot V^\sigma-\alpha\cdot C^\sigma
 \\
 & \textstyle \mbox{s.t.}\textstyle  \sum_{l \in \A}
 \lambda^l \cdot
 \x_i^l = \alpha  \cdot\x_i^\sigma (\b,\bv) , \quad \forall i\in \I,\\
 & ~~~~\textstyle \sum_{l \in \A}
 \lambda^l \cdot
 z_j^l  = \beta  \cdot  z_j^\sigma (\b,\bv), \quad \forall j\in \J,
 \end{aligned}
 \end{equation*}
 \item
 Payment Rule $\{\p^{l}   (\b,\bv),\pv^{l}   (\b,\bv)\}$:

\begin{gather}
 \textstyle
 p_i^{l} (\b,\bv) = \alpha \cdot p_i^\sigma(\b,\bv) \cdot \frac{C_i(\x_i^l)}{ \sum_{l' \in \A}
 \lambda_{l'} \cdot
 C_i(\x_i^{l'}) },\forall i\in\I,\notag\\
 \textstyle
 {q}_j^{l} (\b,\bv) = \beta \cdot q_j^\sigma (\b,\bv) \cdot \frac{V_j(z_j^l)}{ \sum_{l' \in \A}
 	\lambda_{l'} \cdot
 	V_j(z_j^{l'}) },\forall j\in\J,\notag
 \end{gather}
 where $q_j^\sigma (\b,\bv)$ is given in (\ref{eq:reserveprice}). The derivations of $\x ^\sigma(\b,\bv )$, $\z ^\sigma(\b,\bv )$, and $p_i^\sigma (\b,\bv)$ are the same as those in Mechanism \ref{mech:2}, and $V^\sigma$, $C^\sigma$, $C_i(\x_i^l)$, and $V_j(z_j^l)$ are the same as those in Mechanism~\ref{mech:3}.
 \end{itemize}
 \end{mechanism}

Next, we show that Mechanism \ref{mech:reserveprice} with the reserve price is truthful in expectation, but may be not optimal in terms of maximizing the total social welfare.

\begin{proposition}[Truthfulness and Efficiency Loss]
 Mechanism \ref{mech:reserveprice} is truthful in expectation, but is not optimal in terms of maximizing the total social welfare.
\end{proposition}

We show the impact of the reserve price on the social efficiency as follows. Given the reserve price, some task owners, i.e., those with task values lower than the minimum payments given in (\ref{eq:reservepricedata}), will decide not to join the auction. Hence, the maximum social welfare may be reduced. Therefore, there is a tradeoff between the social efficiency and the budget balance. A larger reserve price may lead to a better budget balance and a worse social efficiency. We will show the impact of the reserve price on the budget balance and the social efficiency via simulations in Section \ref{sec:simulation}.

 \section{Simulation Results}\label{sec:simulation}

In this section, we provide simulation results to evaluate the performances of our proposed mechanisms. 
In particular, we first illustrate the performance of our proposed Mechanism \ref{mech:3}. 
Then, we evaluate the performance gain due to data reuse.
Finally, we show 
the impact of the reserve price on the achieved social welfare and the budget balance.

\begin{figure*}
\begin{minipage}[t]{0.32\textwidth}
\centering
\includegraphics[scale=0.32]{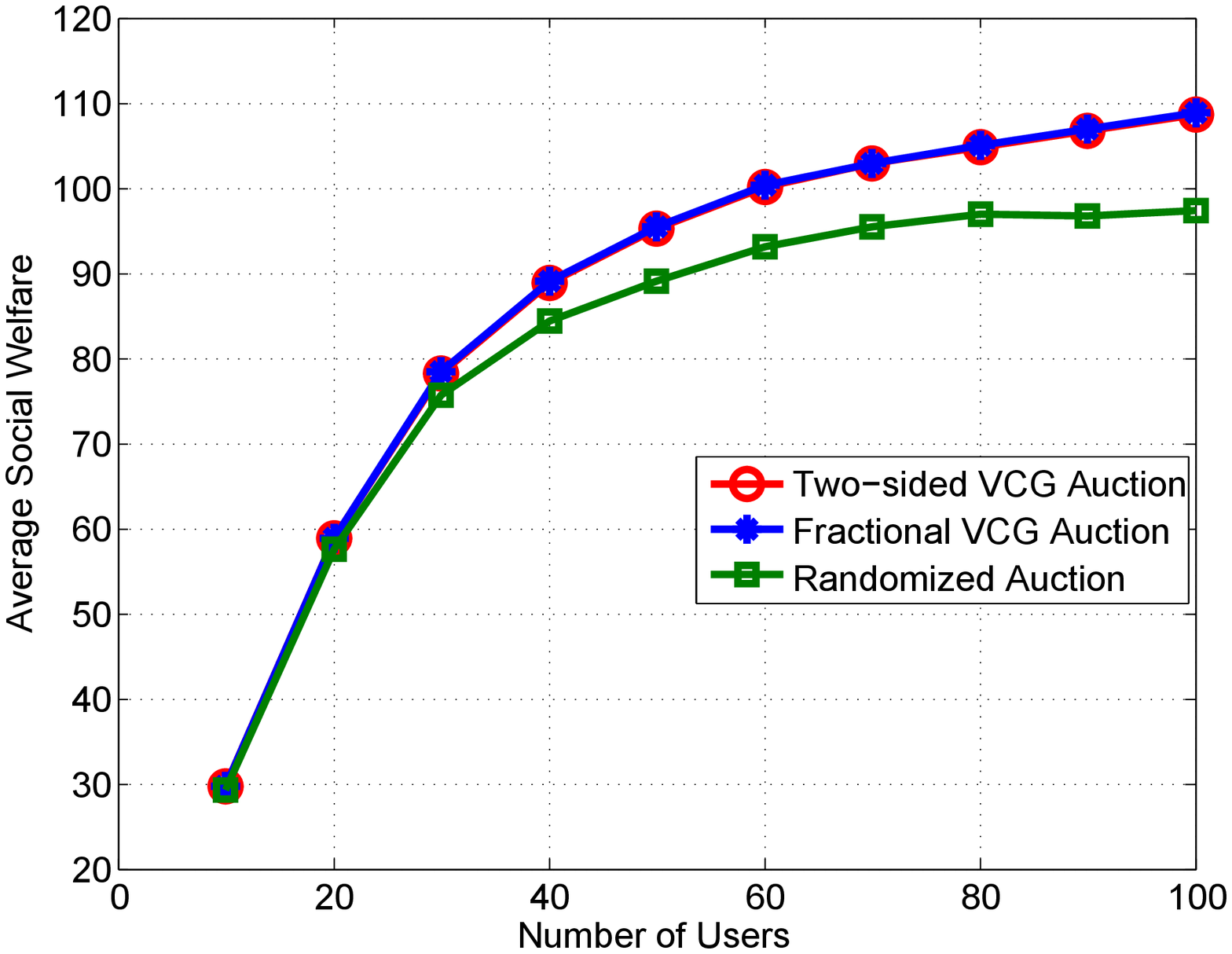}
\caption{Social Welfare vs. Number of Users in Different Auctions. The number of tasks is $50$, the number of data is $30$, and the parameter  $\mu= 1$ in Zipf distribution.}\label{Fig:xx1}
\vspace{-3mm}
\end{minipage}%
\begin{minipage}[t]{0.02\textwidth}
~~~
\end{minipage}%
\begin{minipage}[t]{0.32\textwidth}
\centering
\includegraphics[scale=0.32]{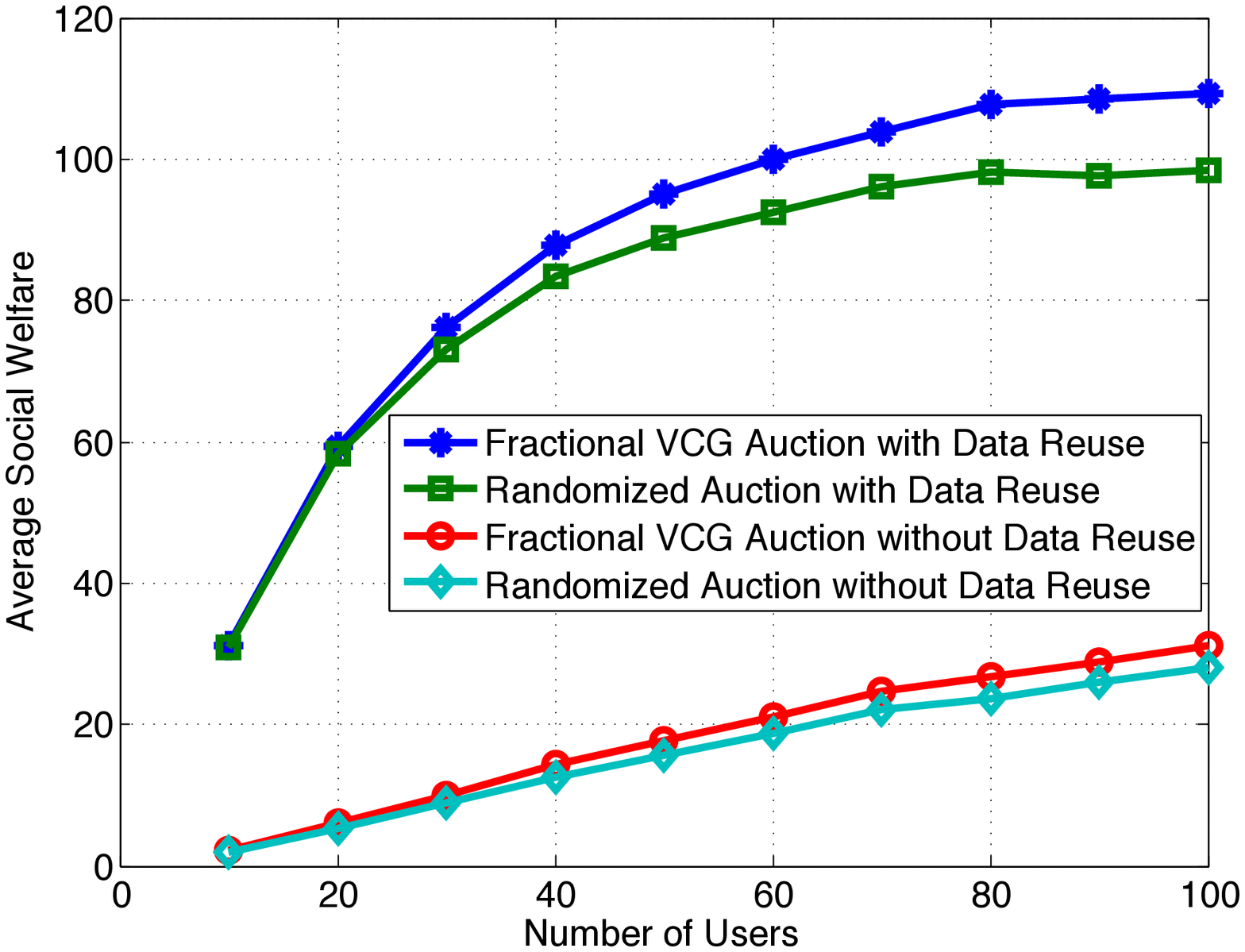}
\caption{Social Welfare vs. Number of Users With/Without Data Reuse. The number of tasks is $50$, the number of data is $30$, and the parameter $\mu= 1$ in Zipf distribution.}\label{Fig:xx2}
\vspace{-3mm}
\end{minipage}%
\begin{minipage}[t]{0.02\textwidth}
~~~
\end{minipage}%
\begin{minipage}[t]{0.32\textwidth}
\centering
\includegraphics[scale=0.32]{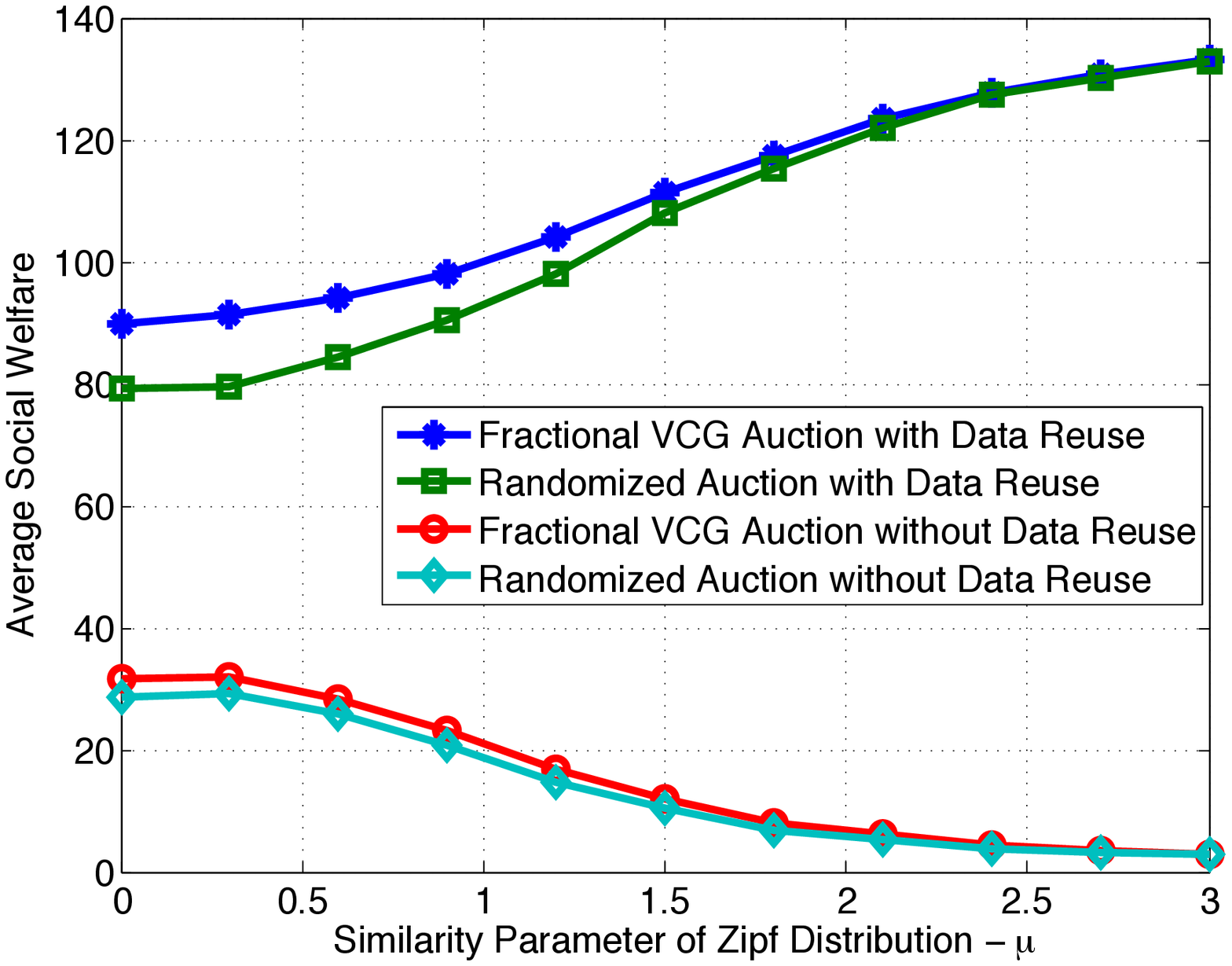}
\caption{Social Welfare vs. Task Similarity With/Without Data Reuse. The number of tasks is $50$, the number of data is $30$, and the number of users is $60$.}\label{Fig:xx3}
\vspace{-3mm}
\end{minipage}%
\end{figure*}

 \begin{figure*}
 \begin{minipage}[t]{0.32\textwidth}
  \centering
 \includegraphics[scale=0.32]{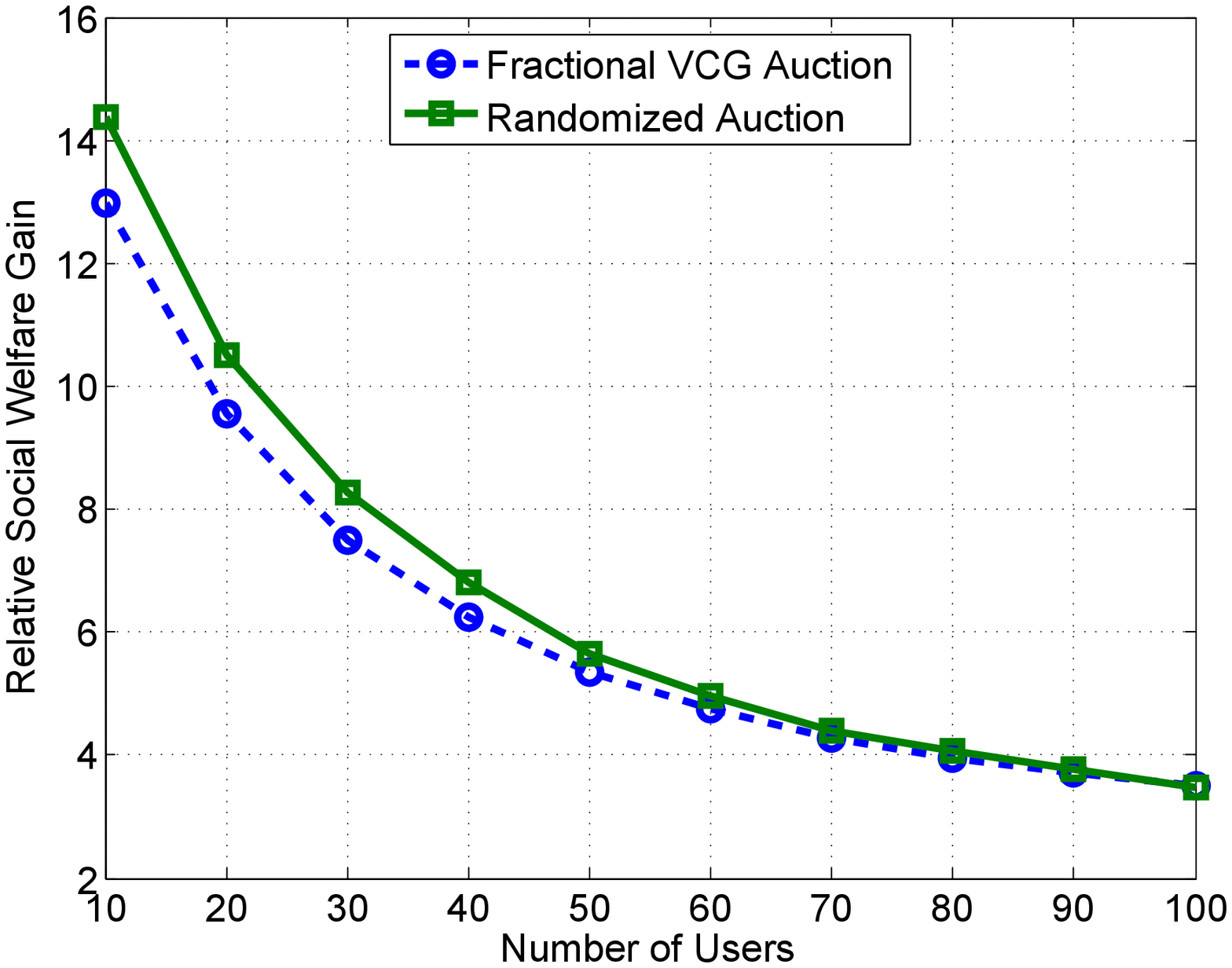}
\caption{Relative Social Welfare Gain vs. Number of Users. The number of tasks is $50$, the number of data is $30$, and the parameter  $\mu= 1$ in Zipf distribution.}\label{Fig:RelativeGaina}
\vspace{-3mm}
  \end{minipage}%
 \begin{minipage}[t]{0.02\textwidth}
 ~~~
 \end{minipage}%
  \begin{minipage}[t]{0.32\textwidth}
   \includegraphics[scale=0.32]{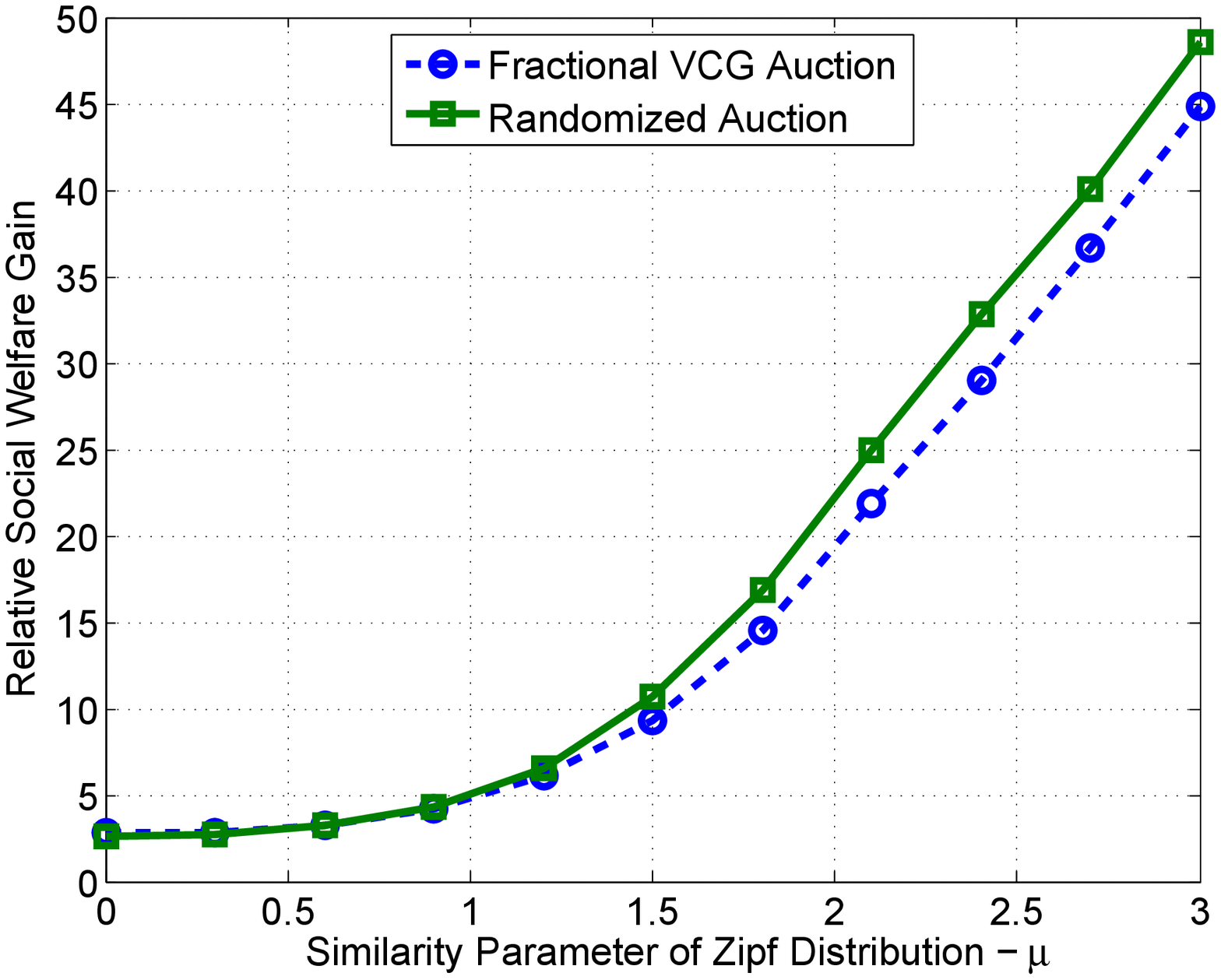}
\caption{Relative Social Welfare Gain vs. Task Similarity. The number of tasks is $50$, the number of data is $30$, and the number of users is 60.}  \label{Fig:RelativeGainb}
\vspace{-3mm}
 \end{minipage}%
 \begin{minipage}[t]{0.02\textwidth}
 ~~~
 \end{minipage}%
   \begin{minipage}[t]{0.32\textwidth}
   \includegraphics[scale=0.32]{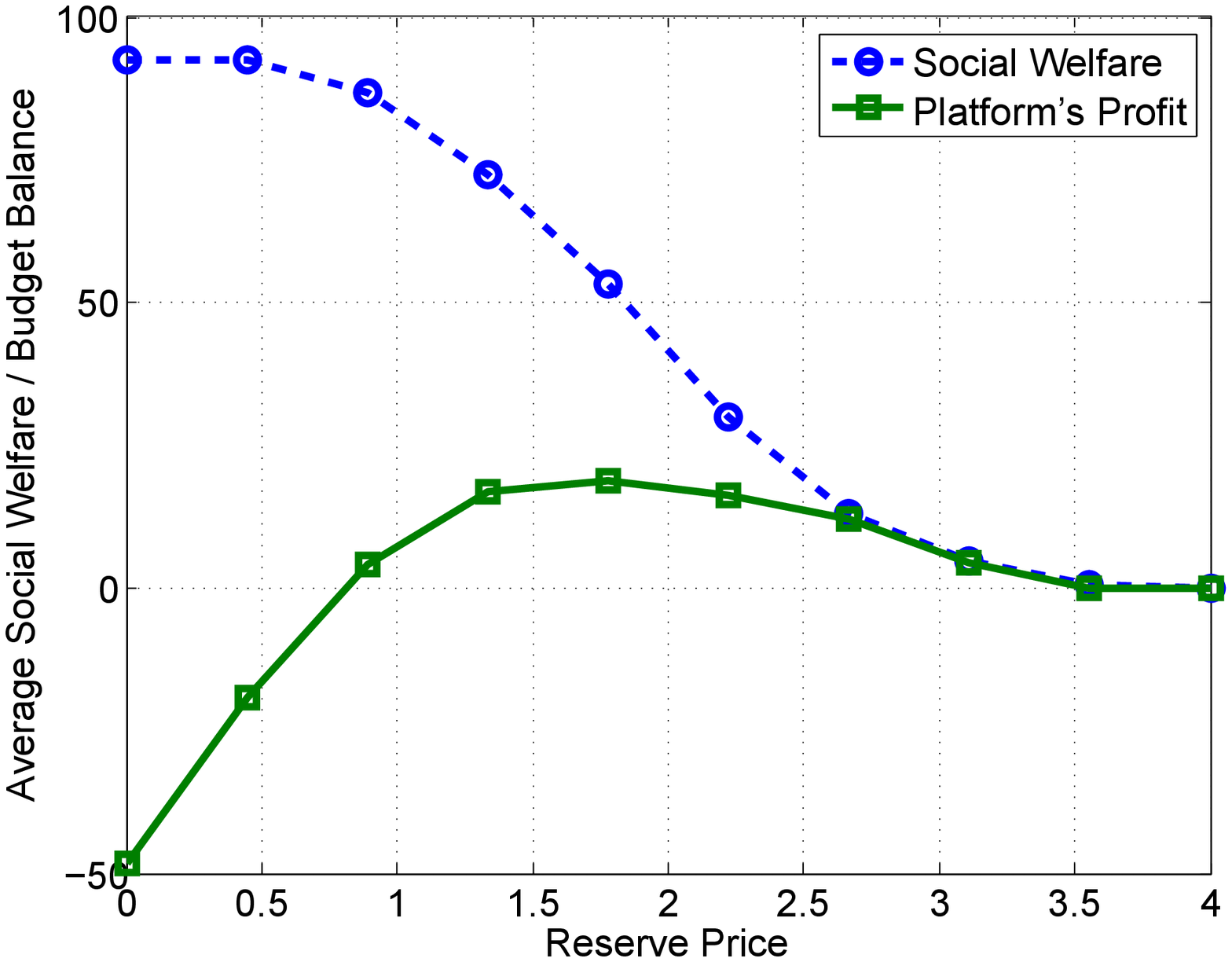}
\caption{Social Welfare and Budget Balance vs. Reserve Price. The number of tasks is $50$, the number of data is $30$, the number of users is $60$, and the parameter $\mu= 1$ in Zipf distribution.}\label{Fig:ReservePricePB}
\vspace{-3mm}
 \end{minipage}%
 \end{figure*}
 
\subsection{Simulation Setup}
In the simulations, we fix the number of tasks to $J=50$ and the number of data items to $K=30$, while varying the number of users from $I = 10$ to $100$ with an increment of $10$.
Each data item is location-based  (such as the temperature at a particular location), and randomly  and uniformly distributed in an area of $1000$m$\times1000$m.
Each user randomly moves to a particular location in a time slot, and can sense all the data items within a distance of $100$m to her location.
The unit cost~$\rho_c$ of each user for sensing \emph{one} data item is chosen randomly   from $[1,5]$, hence the cost for sensing a set $\S$ of data items is $\rho_c \cdot |\S|$.
The unit value $\rho_v$ of each task for \emph{one} data item is also chosen randomly from $[1,5]$, hence the value of a task requiring a set $\S$ of data items is $\rho_v \cdot |\S|$.

We characterize the \emph{task similarity} (in terms of data requirements) in the following way.
First, we define the \emph{popularity} of a data item as the probability that a task requires this particular data item, and denote $P_w $ as the $w$-th highest popularity of all data items.
As demonstrated in \cite{Zipf, Zipf2}, the popularity of data, i.e., $\{P_w, w\in\K\}$, follows a Zipf distribution \cite{zipf-dis} with the p.m.f.
\begin{equation}
P_w=   \frac{(1/w)^\mu}{\sum_{t=1}^{K}(1/t)^{\mu}},\quad \forall w \in \K,
\end{equation}
where $\mu \geq 0$ is the parameter of Zipf distribution.
Obviously, with a larger $\mu$, tasks are more likely to require a small set of high popularity data items (hence with a higher task similarity).
In our simulations, we vary $\mu$ from $0$ to $3$ with an increment of $0.3$.

In each simulation, we choose a particular user number~$I$ and parameter $\mu$, and randomly generate 1000 systems (in terms of tasks' data requirements and unit values  as well as users' sensing capabilities and unit costs) and compute the
average outcome of all systems as the simulation result. 

\subsection{Social Welfare Gap}\label{subsecSWG}

We first compare the social welfare achieved in Mechanisms \ref{mech:1}, \ref{mech:2}, and \ref{mech:3}.
This can help us understand the performance gap of our proposed Mechanism \ref{mech:3} to the maximum social welfare (achieved in Mechanism \ref{mech:1}) or the fractional maximum social welfare (achieved in Mechanism \ref{mech:2}).

Fig. \ref{Fig:xx1} illustrates the average social welfare achieved in different auction mechanisms, under different numbers of users, where the parameter of Zipf distribution is fixed at $\mu=1$.
The red curve (with marker $\circ$) denotes the social welfare achieved in Mechanism \ref{mech:1}, which is equivalent to the maximum social welfare benchmark.
The blue curve (with marker $\ast$) denotes the social welfare achieved in Mechanism~\ref{mech:2}, which is equivalent to the fractional maximum social welfare.
The green curve (with marker \scalebox{0.6}{$\square$}) denotes the social welfare achieved in Mechanism \ref{mech:3}.

From Fig. \ref{Fig:xx1}, we can see that the difference between the maximum social welfare and the fractional maximum social welfare is negligible. Moreover,
the achieved social welfare in all three auction mechanisms increase with the number of users.
The performance gap of the randomized auction in Mechanism \ref{mech:3} to the maximum social welfare (benchmark) increases with the number of users,
and the maximal gap in our simulations (when there are $100$ users) is less than $10\%$.

\subsection{Performance Gain of Data Reuse}\label{subsecGDR}

We now evaluate the performance gain achieved by the data reuse across tasks, by comparing the social welfare achieved in the systems with data reuse and without data reuse.

Next, we will implement the randomized auction (Mechanism \ref{mech:3}) and the fractional VCG auction (Mechanism \ref{mech:2}), by solving Problem P1 and the problem without data reuse (see Appendix A), respectively. We will compare the performance gain due to data reuse in the two mechanisms.

\subsubsection{Impact of the Number of Users}

We first show the impact of the number of users on the performance gain.
Fig. \ref{Fig:xx2} illustrates the achieved social welfare with and without data reuse, under different numbers of users, where the parameter of Zipf distribution is fixed at $\mu=1$.
The blue curve (with marker $\ast$) denotes the social welfare achieved in Mechanism \ref{mech:2} with data reuse, which represents the  maximum social welfare with data reuse based on the observation from Fig. \ref{Fig:xx1}.
The green curve (with marker \scalebox{0.6}{$\square$}) denotes the social welfare achieved in Mechanism \ref{mech:3} with data reuse.
The red curve (with marker $\circ$) and the cyan curve (with marker $\Diamond$) denote the corresponding results without data reuse.

From Fig. \ref{Fig:xx2}, we can see that
the achieved social welfare with and without data reuse both increase with the number of users; and  the increase rate is higher with data reuse, especially when the number of users is small.
Furthermore, with data reuse, the maximum social welfare (benchmark) can   increase up to $350\%$, and the social welfare achieved by the randomized auction (Mechanism \ref{mech:3}) can   increase  up to $300\%$, comparing with those without data reuse.

\subsubsection{Impact of the Task Similarity}

We next show the impact of the task similarity on the performance gain.
Recall that $\mu$ of Zipf reflects the task similarity: \emph{a larger $\mu$ implies a higher task similarity}.
Fig. \ref{Fig:xx3} illustrates the achieved social welfare with and without data reuse, under different values of $\mu$, where the user number is fixed at $I = 60$.
The blue curve (with marker $\ast$) denotes the social welfare achieved in Mechanism \ref{mech:2} with data reuse, which represents the  maximum social welfare with data reuse based on the observation from Fig. \ref{Fig:xx1}.
The green curve (with marker \scalebox{0.6}{$\square$}) denotes the social welfare achieved in Mechanism~\ref{mech:3} with data reuse.
The red curve (with marker $\circ$) and the cyan curve (with marker $\Diamond$) denote the corresponding results without data reuse.

From Fig. \ref{Fig:xx3}, we can see that
the achieved social welfare increases with the task similarity parameter  $\mu$ with data reuse,
while decreases with the parameter $\mu$ without data reuse.
The reason is as follows. With a higher task similarity~$\mu$, most of the tasks' data requirements will concentrate on a smaller set of high popularity data. Hence, with data reuse, a smaller set of users (covering the high popularity data) are needed to cover all the required data requirements, leading to a higher social welfare; while without data reuse, a larger set of users are needed to cover all the required data multiple times, leading to a lower social welfare. Intuitively, without data reuse, the number of ``effective'' users in the high task similarity (i.e., those can sense high popularity data only) is fewer than that in the low task similarity (i.e., those who can sense any data item), hence the social welfare becomes smaller with a high task similarity.

From Fig. \ref{Fig:xx3}, we can further see that
with data reuse, the social welfare (both the maximum social welfare benchmark and the social welfare achieved by the randomized auction in Mechanism \ref{mech:3}) can increase from $300\%$ to $1300\%$ when the task similarity increases from
$\mu= 0$ to $\mu = 3$, comparing with those without data reuse.

Furthermore, Fig. \ref{Fig:RelativeGaina} shows the relative social welfare gain vs. the number of users for Mechanisms \ref{mech:2} and \ref{mech:3}, and  Fig. \ref{Fig:RelativeGainb} shows the relative social welfare gain vs. the task similarity for Mechanisms \ref{mech:2} and \ref{mech:3}. We can see that Fig. \ref{Fig:RelativeGaina} is similar to the green dash line in Fig. \ref{Fig:MonteCarlo}. The relative social welfare gain decreases with the number of users, and increases with the task similarity. In both Figs. \ref{Fig:RelativeGaina} and \ref{Fig:RelativeGainb},  the randomized auction in Mechanism \ref{mech:3} leads to a relative performance gain very close to that of the social optimality (i.e., the fractional VCG auction in Mechanism \ref{mech:2}). This verifies the effectiveness of the proposed randomized auction in Mechanism \ref{mech:3}.

\subsection{Impact of the Reserve Price on the Budget Balance}
Fig. \ref{Fig:ReservePricePB} illustrates  the impact of the reserve price on the social welfare and the budget balance of Mechanism \ref{mech:reserveprice}. We can see that the social welfare always decreases with the reserve price, as a larger reserve price will drive more task owners out of the auction. The MCS platform can achieve the budget balance and gain a positive profit by setting a proper medium value reserve price. Moreover, the platform's profit first increases with the reserve price, due to the increase of the payments from task owners. When the reserve price is high enough, the platform's profit decreases with the reserve price until reaching zero. This is because a high reserve price may drive many task owners out of the auction, leading to a small  social welfare and a small platform's profit.
 
\section{Conclusion}\label{sec:conclusion}
In this work, we proposed a novel three-layer data-centric mobile crowdsensing model, which enables data reuse and leverages both the task similarity and the user heterogeneity.
We focused on the joint task selection and user scheduling problem, aiming at maximizing the social welfare. This problem is NP-hard and is challenging to solve due to the two-sided information asymmetry of selfish task owners and users. To understand the performance gain due to data reuse, we theoretically analyzed the social welfare gain with known statistical information, and proved the bound of the relative performance gain.
To address both the limited computation and incomplete information issues,
we proposed a two-sided randomized auction mechanism, which
is computationally efficient, individually rational, and incentive
compatible (truthful) in expectation. We further proposed a budget-balanced randomized auction mechanism to ensure the profitability of the platform in realistic settings.
Simulations show that the performance gain of data reuse in the randomized auction increases from $300\%$ to $1300\%$ with the increasing of the task similarity. The proposed randomized auction can achieve at least $90\%$ of the maximum social welfare. Furthermore, by choosing proper reserve prices, the randomized auction mechanism can achieve the budget balance without compromise on the truthfulness, individual rationality, and computational efficiency.

\appendices
\section{Social Welfare Maximization without Data Reuse}\label{appendixA}
We now formulate the social welfare maximization problem of the joint task selection and user scheduling without data reuse. In a system without data reuse, one data item provided by a user can only be used by one task. Hence, the total times of the data item $k$ that is required by all selected tasks should be at least sensed by all selected users that contain the data item $k$. In particular,
for each data item $k\in \K$, if it is required by $M$ tasks, then it should be sensed by at least $M$ users, i.e.,
\setcounter{equation}{27}
\begin{equation}\label{eq:NoDataReuse}
\sum_{j\in\J} \mathbf{1}_{(k\in\K_j)}\cdot z_j
\leq
\sum_{i\in\I} \sum_{\S\subseteq\S_i}^{}\mathbf{1}_{(k\in\S)} \cdot x_{i}(\S),
\ \ \forall k\in \K,
\end{equation}
where the indicator $\mathbf{1}_{(x)}=1 $ if $x$ is true, and 0 otherwise. The left-hand side of (\ref{eq:NoDataReuse}) denotes the number of selected tasks requiring data item $k$, and the right-hand side of (\ref{eq:NoDataReuse})  denotes the number of users scheduled for sensing data item~$k$.

Formally, we formulate the social welfare maximization problem of the joint task selection and user scheduling without data reuse as follows.
$$
\begin{aligned}
& & \max_{\boldsymbol{x},\boldsymbol{z}} & ~~~ V(\boldsymbol{z}) - C(\boldsymbol{x})
\\
& & \mbox{s.t.} & ~~~ (2), (3), (4), \eqref{eq:NoDataReuse}, \quad \forall i \in \I, j \in \J, k \in \K;
\\
& & \mbox{var.} & ~~~  x_i(\S) \in \{0, 1\},\quad \forall \S\subseteq\S_i , i \in \I ;
\\
& & & ~~~   z_j \in \{0, 1\},\quad \forall j \in \J.
\end{aligned}
$$

Let the optimal solution to the above problem be $(\bar{\x}^o,\bar{\z}^o)$,  and the optimal solution to Problem P1 be $(\x^o,\z^o)$. Then the relative social welfare (performance) gain is
$$
\gamma=\frac{V(\z^o) - C(\x^o)}{V(\bar{\z}^o) - C(\bar{\x}^o)}.
$$

\section{Proof of Proposition 1}
\begin{proof}
By transforming the domains of integration $\T_m (m=1,\cdots,\min\{I,J\})$, we can derive the social welfare without data reuse as:
\begin{align}
& SW_{n} =\sum_{m=1}^{\min\{I,J\}-1}SW_{n}[m]+SW_{n}[\min\{I,J\}]\notag\\
&=\sum_{m=1}^{\min\{I,J\}}\int_{v_{m:J}\geq c_{m:I}}\notag\\
&~~~~~~(v_{m:J}-c_{m:I})[(\min\{I,J\})!]^2\prod_{k=1}^{\min\{I,J\}}f(v_k)g(c_k)\d c\d v\notag\\
&=\sum_{m=1}^{\min\{I,J\}}\int_{0}^{1}\int_{0}^{v_{1:J}}\cdots\int_{0}^{v_{m-1:J}}\cdots\int_{0}^{v_{\min\{I,J\}-1:J}}\notag\\
&~~\int_{0}^{v_{m:J}}\int_{c_{1:I}}^{v_{2:J}}\cdots\int_{c_{m-1:I}}^{v_{m:J}}\int_{c_{m:I}}^{1}\int_{c_{m+1:I}}^{1}\cdots\int_{c_{\min\{I,J\}-1:I}}^{1}\notag\\
&~~(v_{m:J}-c_{m:I})[(\min\{I,J\})!]^2\prod_{k=1}^{\min\{I,J\}}f(v_k)g(c_k)\notag\\
&~~\d c_{\min\{I,J\}:I}\cdots\d c_{2:I}\d c_{1:I}\d v_{\min\{I,J\}:J}\cdots\d v_{2:J}\d v_{1:J}.\notag
\end{align}
In particular, if  ($v_j,j\in\J)$ and $(c_i,i\in\I)$ follow i.i.d. uniform distributions, the integral can be further simplified since $\prod_{k=1}^{\min\{I,J\}}f(v_k)g(c_k)=1$. We can obtain the limiting values of $ SW_{n}$ under different numbers of tasks and users, respectively. Recall that, without data reuse, we sort the task values in the descending order, i.e., $v_{1:J}\geq v_{2:J}\geq \cdots \geq v_{J:J}$, and sort sensing costs in the ascending order, i.e., $c_{1:I}\leq c_{2:I}\leq \cdots \leq c_{I:I}$. Then, there is a threshold $m,1\leq m\leq\min\{I,J\}$, such that the $m$-th task value is no greater than the $m$-th user cost. The social welfare maximization selection selects tasks with values $v_{1:J},\cdots, v_{m:J}$ and users with sensing costs $c_{1:I},\cdots, c_{m:I}$. In particular, we have the following results.
\begin{itemize}
\item If the number of users $I$ and the number of tasks $J$ are identical and sufficiently large, i.e., $I=J\to\infty$, then this equivalently means that half of the $J$ tasks with the task values uniformly distributed in $[0.5,1]$ and half of the $I$ users with the sensing costs uniformly distributed in $[0,0.5]$ are matched. That is, the threshold $m=J/2=I/2$.  Furthermore, the mean task value of the selected tasks is $0.75$, and the the mean sensing cost of the selected users is $0.25$. In this case, the social welfare without data reuse is 
$ SW_{n}=0.75\cdot J/2-0.25\cdot I/2=J/4.$
\item If the  number of users~$I$ is sufficiently large, i.e., $I\to\infty$ with a limited $J$ of tasks, then the threshold $m=J$. That is, all $J$ tasks with the task values uniformly distributed in $[0,1]$ and $J$ users with negligible costs are matched and selected. The social welfare without data reuse is 
$ SW_{n}=0.5\cdot J-0\cdot J=J/2.$
\item If the  number of tasks~$J$ is sufficiently large, i.e., $J\to\infty$ with a limited~$I$ of users, then the threshold $m=I$. That is, $I$ tasks with the task values 1 and all $I$ users with the sensing costs uniformly distributed in $[0,1]$ are matched and selected. The social welfare without data reuse is 
$SW_{n}=1\cdot I-0.5\cdot I=I/2.$
\end{itemize}
The three limiting results thus provide the upperbounds of the social welfare without data reuse, respectively. This completes the proof.
\end{proof}

\section{Proofs of Proposition 2 and 3}
\begin{proof}

By transforming the domain of integration $\R$, we can derive the social welfare with data reuse as:
\begin{align}
& SW_{r} =\int_{\R}(v-c)f_{\sum_{j=1}^{J}v_j}(v)g_{\min\{c_i,i\in\I\}}(c)\mathrm{d}c\mathrm{d}v\notag\\
&=\int_{0}^{1}\int_{0}^{v}(v-c)f_{\sum_{j\in\J}^{}v_j}(v)g_{\min\{c_i,i\in\I\}}(c)\mathrm{d}c\mathrm{d}v\notag\\
&+\sum_{j=1}^{J-1}\int_{j}^{j+1}\!\!\!\int_{0}^{1}(v-c)f_{\sum_{j\in\J}^{}v_j}(v)g_{\min\{c_i,i\in\I\}}(c)\mathrm{d}c\mathrm{d}v.\notag
\end{align}
In particular, if  ($v_j,j\in\J)$ and $(c_i,i\in\I)$ follow i.i.d. uniform distributions, then the term $\sum_{j\in\J}^{}v_j$ follows the Irwin-Hall distribution.
It turns out that the Irwin-Hall distribution is complicated. To simplify the analysis, we can use the \emph{Normal Distribution} to approximate the Irwin-Hall distribution. The approximation is increasingly better when $J$ increases. In particular, the term $\sum_{j\in\J}^{}v_j$ follows a normal distribution with the mean $J/2$ and the standard deviation $\sqrt{\frac{J}{12}}$. 

Furthermore, it is easy to show that the term $\min\{c_i,i\in\I\}$ follows a distribution with the p.d.f.
$$
g_{\min\{c_i,i\in\I\}}(c)=I(1-c)^{I-1}.
$$

Based on the above distributions, we can compute the social welfare with data reuse as
$$
\begin{aligned}
SW_r&=\int_{c=0}^{1}\int_{v=c}^{\infty}(v-c)f_{\sum_{j\in\J}^{}v_j}(v)g_{\min\{c_i,i\in\I\}}(c)\d v\d c\\
&=\int_{c=0}^{1}\int_{v=c}^{\infty}(v-c)I(1-c)^{I-1}f_{\sum_{j\in\J}^{}v_j}(v)\d v\d c\\
&{\geq} \int_{c=0}^{1}I(1-c)^{I-1}\int_{-\infty}^{\infty}(v-c)f_{\sum_{j\in\J}^{}v_j}(v)\d v\d c\\
&=\int_{c=0}^{1}I(1-c)^{I-1}\left(\frac{J}{2}-c\right)\d c\\
&=\frac{J}{2}-1+\frac{I}{I+1}.
\end{aligned}
$$
We have obtained the lower bound of the social welfare with data reuse. We can see that the social welfare with data reuse is $J/2$ when the number of users $I$ is large enough. Intuitively, in this case, all $J$ tasks are selected with a negligible sensing cost. 

Furthermore, together with Proposition 1, we have the following results. First, the lower bound  of $\gamma$ is $(J/2)/(J/4)=2, $ when $I=J\to\infty$. Second, the lower bound  of $\gamma$ is $(J/2)/(J/2)=1,$ when $I\to\infty$, with a limited $J$. Finally, the lower bound  of $\gamma$ is  $(J/2)/(I/2)=J/I,$ when $J\to\infty$, with a limited $I$. This completes the proofs.
\end{proof}

\section{Proofs of Propositions 4 and 5}
\begin{proof}
The  two-sided VCG auction mechanism $\Omega^o$ in Mechanism 1 is designed based on the classic VCG mechanism. The classic VCG mechanism is the well-known mechanism that maximizes the social welfare (efficient), while keeping incentive compatibility
(truthfulness). Moreover, the classic VCG mechanism is also individually rational, as it ensures a non-negative payoff for each winning bidder. We prove the detailed results for $\Omega^o$ in the following.

(i). Social Welfare Maximization (SWM). This follows immediately from the fact that task owners and users bid their costs truthfully in the mechanism $\Omega^o$, and that it implements an efficient outcome with respect to these bids. The truthful bidding will be proved in the following incentive-compatible property.

(ii). Incentive Compatibility (IC). Let $W^o(\b,\bv)$ be the optimal social welfare achieved in $\Omega^o$ with users' bids $\b$ and task owners' bids $\bv$. Consider an arbitrary user $i$ with the bid $\b_i$. We fix the bids of the other users to be $\boldsymbol{b}_{-i}$ and fix the bids of task owners to be $\bv$. Let $W_{-i}^o(\emptyset,\boldsymbol{b}_{-i},\bv)$ be the optimal social welfare among bidders $\{\I\setminus\{i\}\}\cup\J$ with respect to $\boldsymbol{b}_{-i}$ and $\bv$. 
Then the utility of user $i$  is the difference between the obtained payment and the sensing cost of user $i$, i.e.,
\begin{align}
& U_i(\b_i,\boldsymbol{b}_{-i},\bv) =p_i^o(\b_i,\boldsymbol{b}_{-i},\bv)- \sum_{\S \subseteq \S_i}  b_i (\S) \cdot x_i^o (\S) \notag\\
&=\sum_{j\in \J}  u_j \cdot z_j^o (\b,\bv )
- \sum_{n \in \I \setminus \{i\}} \sum_{\S \subseteq \S_n}  b_n (\S) \cdot x_n^o (\S)\notag\\
&~~~~~~~~~~~~~~~~~~~- W^o_{-i}(\emptyset,\boldsymbol{b}_{-i},\bv) -\sum_{\S \subseteq \S_i}  b_i (\S) \cdot x_i^o (\S) \notag\\
&=\sum_{j\in \J}  u_j \cdot z_j^o (\b,\bv )
- \sum_{n \in \I} \sum_{\S \subseteq \S_n}  b_n (\S) \cdot x_n^o (\S)\notag\\
&~~~~~~~~~~~~~~~~~~~~~~~~~~~~~~~~~~~~~~~~~~~~~~~~~~- W^o_{-i}(\emptyset,\boldsymbol{b}_{-i},\bv)\notag\\
&=W^o(\b_i,\boldsymbol{b}_{-i},\bv)-W^o_{-i}(\emptyset,\boldsymbol{b}_{-i},\bv).\notag
\end{align}
Similarly, the utility of user $i$ with the bid $\boldsymbol{b}'_i$ is
\begin{align}
& U_i(\boldsymbol{b}'_i,\boldsymbol{b}_{-i},\bv)=W^o(\b'_i,\boldsymbol{b}_{-i},\bv)-W^o_{-i}(\emptyset,\boldsymbol{b}_{-i},\bv).\notag
\end{align}
By reporting the true cost $\boldsymbol{b}_i=\boldsymbol{c}_i$, we have
\begin{align}
& U_i(\c_i,\boldsymbol{b}_{-i},\bv)-U_i(\boldsymbol{b}'_i,\boldsymbol{b}_{-i},\bv)\notag\\
&=W^o(\c_i,\boldsymbol{b}_{-i},\bv)-W^o(\b'_i,\boldsymbol{b}_{-i},\bv)\geq 0. \label{eq:xx1}
\end{align}
The inequality here follows because $W^o(\c_i,\b_{-i},\bv)$ is the optimal social welfare with respect to the true profile of costs $(\boldsymbol{c}_i,i\in\I)$. From (\ref{eq:xx1}), we can see that reporting the true sensing cost is a dominant strategy for each user $i\in\I$.

 Similarly, we can show that the utility of task owner $j$  is
$$
\begin{aligned}
& U_j(\b,v_j,\boldsymbol{u}_{-j})-U_j(\b,u'_j,\boldsymbol{u}_{-j})\\
& =W^o(\b,v_j,\boldsymbol{u}_{-j})-W^o(\b,u'_j,\boldsymbol{u}_{-j})\geq 0,
\end{aligned}
$$
The inequality follows because $W^o(\b,v_j,\boldsymbol{u}_{-j})$ is the optimal social welfare with respect to the true profile of task values $(v_j,j\in\J)$. Furthermore, truthful reporting is a dominant strategy for each task owner $j\in\J$.

This completes the proof of (IC), because the utilities of user $i$ and task owner $j$ do not depend on her own reported cost and task value, respectively.

(iii). Individual Rationality (IR). Given the other users' bids $\boldsymbol{b}_{-i}$ and task owners' bids $\bv$, the utility of user $i$ with the bid $\boldsymbol{b}_i=\c_i$ is
$$
U_i(\c_i,\boldsymbol{b}_{-i},\bv)=W^o(\c_i,\boldsymbol{b}_{-i},\bv)-W^o_{-i}(\emptyset,\boldsymbol{b}_{-i},\bv).
$$
We next show that $U_i(\c_i,\boldsymbol{b}_{-i},\bv)$ is non-negative. Recall that $W^o_{-i}(\emptyset,\boldsymbol{b}_{-i},\bv)$ is the optimal social welfare of an efficient allocation $(\x'(\b,\bv ),\z'(\b,\bv ))$ under the case where we give $\emptyset$ to user~$i$. By definition, the total social welfare $W^o$ of the efficient allocation $(\x ^o(\b,\bv ),\z ^o(\b,\bv ))$ under the profiles $\b_i=\boldsymbol{c}_i$ and $u_j=v_j$ is at least that of any other feasible allocation, in particular, the one $(\x'(\b,\bv ),\z'(\b,\bv ))$ with $\emptyset$ to user $i$.
Thus we have
\begin{align}
W^o(\c_i,\boldsymbol{b}_{-i},\bv)&\geq W^o(\emptyset,\boldsymbol{b}_{-i},\bv)=W^o_{-i}(\emptyset,\boldsymbol{b}_{-i},\bv).\notag
\end{align}
Similarly, for task owner $j$, let $W^o_{-j}(\b,\emptyset,\bv_{-j})$ be the optimal social welfare among bidders $\I\cup\{\J\setminus \{j\}\}$ with respect to $\boldsymbol{b}$ and $\bv_{-j}$, we have 
$$
U_j(\boldsymbol{b},v_j,\bv_{-j})=W^o(\b,v_j,\boldsymbol{u}_{-j})-W^o_{-j}(\b,\emptyset,\bv_{-j})\geq 0.
$$
This completes the proof of (IR). Furthermore, the proof of  Proposition 5 is similarly to that of Proposition 4, except that the allocation rules in terms of $\x$ and $\z$ are in the fractional domain. We thus skip the detailed proof.
\end{proof}

\section{Proof of Proposition 6}
\begin{proof}
Let $W^\ast(\b,\bv)$ be the optimal social welfare achieved in the fractional VCG auction mechanism~$\Omega^{\ast}$ with users' bids $\b$ and task owners' bids $\bv$. Let $W_{-i}^\ast(\emptyset,\boldsymbol{b}_{-i},\bv)$ be the optimal social welfare among bidders $\{\I\setminus\{i\}\}\cup\J$ with respect to $\boldsymbol{b}_{-i}$ and $\bv$. Given $\Omega^{\ast}$, the utility of user $i$ is
\begin{align}
& U_i(\boldsymbol{b}_i,\boldsymbol{b}_{-i},\bv) = p_i^{\ast}(\boldsymbol{b},\bv)- \sum_{\S \subseteq \S_i}  b_i (\S) \cdot x_i^{\ast} (\S) \notag\\
&=\sum_{j\in \J}  u_j \cdot z_j^{\ast} (\b,\bv )
- \sum_{n \in \I \setminus \{i\}} \sum_{\S \subseteq \S_n}  b_n (\S) \cdot x_n^{\ast} (\S)\notag\\
&~~~~~~~~~~~~~~~~- W^{\ast}_{-i}(\emptyset,\boldsymbol{b}_{-i},\bv)- \sum_{\S \subseteq \S_i}  b_i (\S) \cdot x_i^{\ast} (\S) \notag\\
&=\sum_{j\in \J}  u_j \cdot z_j^{\ast} (\b,\bv )
- \sum_{n \in \I} \sum_{\S \subseteq \S_n}  b_n (\S) \cdot x_n^{\ast} (\S)\notag\\
&~~~~~~~~~~~~~~~~~~~~~~~~~~~~~~~~~~~~~~~~~~~~~ - W^{\ast}_{-i}(\emptyset,\boldsymbol{b}_{-i},\bv)\notag\\
&=W^{\ast}(\boldsymbol{b}_i,\boldsymbol{b}_{-i},\bv)- W^{\ast}_{-i}(\emptyset,\boldsymbol{b}_{-i},\bv).\notag
\end{align}
The utility of task owner $j$ is the difference between the task value and the payment of task owner $j$. That is,
\begin{align}
& U_j(\boldsymbol{b},u_j,\bv_{-j}) =u_j \cdot z_j^{\ast} (\b,\bv )-q_j^{\ast}(\boldsymbol{b},\bv)\notag\\
&=u_j \cdot z_j^{\ast} (\b,\bv )-W^\ast_{-j}(\boldsymbol{b},\emptyset,\bv_{-j})\notag\\
&~~~~~~~~~~~+\sum\limits_{j\in \J \setminus \{j\}} u_j z_j^\ast (\b,\bv )-\sum\limits_{n \in \I } \sum\limits_{\S \subseteq \S_n}  b_n (\S) x_n^\ast (\S) \notag\\
&=\sum_{j\in \J}  u_j \cdot z_j^{\ast} (\b,\bv )
- \sum_{n \in \I } \sum_{\S \subseteq \S_n}  b_n (\S) \cdot x_n^{\ast} (\S)\notag\\
&~~~~~~~~~~~~~~~~~~~~~~~~~~~~~~~~~~~~~~~~~~~~~~~- W^\ast_{-j}(\boldsymbol{b},\emptyset,\bv_{-j})\notag\\
&=W^{\ast}(\boldsymbol{b},u_j,\bv_{-j})-W^\ast_{-j}(\boldsymbol{b},\emptyset,\bv_{-j}).\notag
\end{align}

We can clearly see that, given the $({\alpha},\beta)$-scaled fractional allocation and payment $\{\x^{\ast}_\alpha(\cdot), \z^{\ast}_\beta(\cdot); \p^{\ast}_\alpha(\cdot),\pv^{\ast}_\beta(\cdot)\}$,  the above utilities $U_i(\boldsymbol{b}_i,\boldsymbol{b}_{-i},\bv)$ and $ U_j(\boldsymbol{b},u_j,\bv_{-j}) $ will be scaled by $\alpha$ and $\beta$, respectively. More intuitively, we can also show that the $({\alpha},\beta)$-scaled fractional mechanism is incentive-compatible because the utility of user $i$ and the utility of task owner $j$ still do not depend on her own reported cost and task value, respectively. This equivalently shows that the  incentive-compatible (truthful) property still holds for the $({\alpha},\beta)$-scaled fractional mechanism.
\end{proof}

\section{Proof of Proposition 7}
\begin{proof}
We now prove the incentive compatibility in expectation for the randomized auction mechanism $\wmech^{\dag}$. According to the  randomized auction mechanism $\wmech^{\dag}$ in Mechanism 3, the payment obtained by user $i$ is
$$
p_i^{l} (\b,\bv) = \alpha \cdot p_i^\ast (\b,\bv) \cdot \frac{C_i(\x_i^l)}{ \sum_{l' \in \A}
\lambda_{l'} \cdot
C_i(\x_i^{l'}) }.
$$
Let $X_i(\S)$ be a binary random variable that takes value 1 if and only if $\S_i=\S$. Given the users' bids $\b=\c$, the random payment obtained by user $i$ is
\begin{align}
& PU_i = \sum_{\S\subseteq\S_i}\frac{c_i(\S)X_i(\S)}{ \sum_{l' \in \A}
\lambda_{l'} \cdot C_i(\x_i^{l'}) }\cdot\alpha\cdot p_i^\ast (\b,\bv)\notag\\
&~~~=\sum_{\S\subseteq\S_i}\frac{c_i(\S)X_i(\S)}{\alpha c_i(\x_i^{\ast}) }\alpha p_i^\ast (\b,\bv)\notag
\end{align}
First notice that the expectation $\mathbb{E}[X_i(\S)]=\mathrm{Pr}[\S_i=\S]=\sum_{l \in \A}
\lambda^l \cdot
\x_i^l=\alpha\cdot\x_i^\ast (\b,\bv)$. Then the expected payment obtained by user $i$ is
\begin{align}
\mathbb{E}[PU_i]&=\mathbb{E}\left[\sum_{\S\subseteq\S_i}\frac{c_i(\S)X_i(\S)}{\alpha c_i(\boldsymbol{x}_i^{\ast})}\alpha p_i^\ast (\b,\bv)\right]\notag\\
&=\sum_{\S\subseteq\S_i}\frac{c_i(\S)\mathbb{E}[X_i(\S)]}{\alpha c_i(\boldsymbol{x}_i^{\ast})}\alpha p_i^\ast (\b,\bv)\notag\\
&=\sum_{\S\subseteq\S_i}\frac{c_i(\S)\alpha x_i^{\ast}(\S)}{\alpha c_i(\boldsymbol{x}_i^{\ast})}\alpha p_i^\ast (\b,\bv)\notag\\
&=\alpha p_i^\ast (\b,\bv).\notag
\end{align}

Furthermore, the random utility of user $i$ is $U_i(\b,\bv)=PU_i-\sum_{\S\subseteq\S_i}b_i(\S)X_i(\S)$. Then we derive the expected utility of user $i$ as
\begin{align}
\mathbb{E}[U_i(\b,\bv)]&=\mathbb{E}\left[PU_i-\sum_{\S\subseteq\S_i}b_i(\S)X_i(\S)\right]\notag\\
&=\mathbb{E}[PU_i]-\sum_{\S\subseteq\S_i}b_i(\S)\mathbb{E}[X_i(\S)]\notag\\
&=\alpha p_i^\ast (\b,\bv)-\alpha\sum_{\S\subseteq\S_i}b_i(\S)x_i^{\ast}(\S)\notag\\
&=\alpha U_i^{\ast}(\b,\bv).\notag
\end{align}
Similarly, we can show that the expected payment charged to task owner $j$ is $\mathbb{E}[PT_j] =\beta q_j^\ast (\b,\bv)$. The expected utility of task owner $j$ is
$$\mathbb{E}[U_j(\b,\bv)]=\beta U_j^{\ast}(\b,\bv).$$

We can see that the expected utilities and the expected payments of users and task owners are scaled by $\alpha$ and $\beta$, respectively. According to Proposition 6, the incentive-compatible property still holds for the $({\alpha},\beta)$-scaled fractional mechanism. This shows that the  randomized auction mechanism $\wmech^{\dag}$ in Mechanism 3 is incentive-compatible (truthful) in expectation. This completes the proof.
\end{proof}

\section{Proof of Proposition 8}
\begin{proof}
Suppose the realization of the randomized auction mechanism $\wmech^\dag$ is $(\x^l,\z^l;\p^l,\pv^l)$. Then, the user scheduling is $(\S_1^l,\cdots,\S_I^l)$ for all users in $\I$. By truth telling, the utility of user $i\in\I$ in the randomized auction mechanism $\wmech^\dag$ is
\begin{align}
U_i(\c_i,\b_{-i},\bv) & =p_i^l(\b,\bv)-c_i(\S_i^l)\notag\\
& =\frac{c_i(\S_i^l)}{\alpha c_i(\x_i^{\ast})}\cdot \alpha \cdot p_i^{\ast}(\b,\bv)-c_i(\S_i^l)\notag\\
&=\frac{c_i(\S_i^l)}{ c_i(\x_i^{\ast})} p_i^{\ast}(\b,\bv)-c_i(\S_i^l)\notag\\
&=\frac{c_i(\S_i^l)}{ c_i(\x_i^{\ast})} (p_i^{\ast}(\b,\bv)-c_i(\x_i^{\ast}))\geq0.\notag
\end{align}
The inequality follows from the individual rationality of the fractional VCG auction $\mech^*$, i.e., $p_i^{\ast}(\b,\bv)-c_i(\x_i^{\ast})\geq0$.

Similarly, by truth telling, the utility of task owner $j\in\J$ in the randomized auction mechanism $\wmech^\dag$ is
\begin{align}
U_j(\b,v_j,\bv_{-j})&=v_jz^l_j-q_j^l(\b,\bv)\notag\\
&=v_jz^l_j-\frac{v_jz^l_j}{\beta v_j(\z_j^{\ast})}\cdot \beta \cdot q_j^{\ast}(\b,\bv),\notag\\
&=v_jz^l_j-\frac{v_jz^l_j}{v_j(\z_j^{\ast})} q_j^{\ast}(\b,\bv),\notag\\
&=\frac{v_jz^l_j}{v_j(\z_j^{\ast})} (v_j(\z_j^{\ast})- q_j^{\ast}(\b,\bv))\geq0.\notag
\end{align}
The inequality follows from the individual rationality of the fractional VCG auction $\mech^*$, i.e., $v_j(\z_j^{\ast})- q_j^{\ast}(\b,\bv)\geq0$.

The above results show that the randomized auction mechanism $\wmech^\dag$ satisfies the individual rationality, given any realizations of the randomized auction mechanism. This further shows that the randomized auction mechanism $\wmech^\dag$ is individually rational, which completes the proof.
\end{proof}

\section{Proofs of Proposition 9 and 10}
\begin{proof}
First, recall that the randomized auction mechanism $\wmech^\dag$ in Mechanism 3 requires the task selections to satisfy
$$
\sum_{j\in \J}  \sum_{l \in \A}
\lambda^l \cdot z_j^l \cdot v_j = \beta  \cdot \sum_{j\in \J} z_j^\ast (\b,\bv) \cdot v_j.
$$
Then it immediately follows that the expected total task value achieved in the randomized auction $\wmech^\dag$ equals a $\beta^*$-fraction of the total task value achieved in the fractional VCG auction $\mech^*$, where $\beta^\ast$ is obtained by optimizing $\beta$. We next show the total sensing cost of users. Similarly, let $X_i(\S)$ be a binary random variable that takes value 1 if and only if $\S_i=\S$, then the expected total sensing cost is given by
$$
\begin{aligned}
\mathbb{E}\left[\sum_{i\in\I}b_i(\S_i^l)\right]&=\mathbb{E}\left[\sum_{i\in\I}\sum_{\S\subseteq\S_i}b_i(\S)X_i(\S)\right]\\
&=\sum_{i\in\I}\sum_{\S\subseteq\S_i}b_i(\S)\mathbb{E}[X_i(\S)]\\
&=\sum_{i\in\I}\sum_{\S\subseteq\S_i}b_i(\S)\cdot \alpha \cdot x_i^{\ast}(\S)\\
&=\alpha \sum_{i\in\I}\sum_{\S\subseteq\S_i}b_i(\S)x_i^{\ast}(\S)
\end{aligned}
$$
This shows that the expected total sensing cost in the randomized auction $\wmech^\dag$ equals an $\alpha^*$-fraction of the total sensing cost in the fractional VCG auction $\mech^*$, where $\alpha^\ast$ is obtained by optimizing $\alpha$.

Furthermore, given the \textsf{DEMO} scheme in (22) and (23), we know that the probability distribution $\lambda^l$ satisfies
$\x^*=\sum_{l\in\A}\lambda^l\boldsymbol{x}^{l}$
with $\alpha=1$, and all $z_j^\ast$ are scaled by the same factor $\beta$, where
$$
\beta=\min\limits_{ j\in\J}\left(\frac{\sum_{l \in \A}\lambda^l \cdot z_j^l}{z_j^\ast}\right).
$$
This shows that Mechanism 3 with the \textsf{DEMO} scheme can achieve an $\alpha^{\ast}$-fraction of total sensing cost in Mechanism~2 with a $\beta^{\ast}$-fraction of the total task value in Mechanism~2. Meanwhile, we have $\alpha^{\ast}=1$ and $\beta^{\ast}=\min\limits_{ j\in\J}\left(\frac{\sum_{l \in \A}\lambda^l \cdot z_j^l}{z_j^\ast}\right).$ This completes the proof.
\end{proof}

\section{Proof of Proposition 11}
\begin{proof}
First notice that the allocation and payment rule in Mechanism 4 have the same structures as those in Mechanism 3. According to Proposition 7, the truthfulness in expectation of Mechanism 4  can be readily derived from the truthfulness of the underlying fractional VCG auction mechanism with the reserve price. Hence, it suffices to show that  the fractional VCG auction with the reserve price is truthful. 

Since the payment rule of users does not change, compared with the fractional VCG auction mechanism $\Omega^{\ast}$, users will still truthfully report their sensing costs according to Proposition 5. We only need to prove the truthful reporting of task owners.

To show the truthful reporting of task owners, we first notice an important observation. That is, for any task owner $j\in\J$, her utility with task value $v_j$ as well as the bid reduction and payment rule in Definition 2 is identical to her utility in Mechanism 2 when she has task value $v_j-\underline{q}_j^\sigma$. In particular, given the task value $v_j$ and the minimum payment $\underline{q}_j^\sigma$, the utility of task owner $j$ from being selected (i.e., $z_j^\sigma=1$) is $v_j-q_j^\sigma(\b,\bv')-\underline{q}_j^\sigma$. Therefore the truthful reporting of task owners in Mechanism 2 with the bids reduction and payment rule in  Definition 2 follows from the truthfulness of Mechanism 2.

We note that Mechanism 2 with the bids reduction and payment rule in Definition 2 is efficient with respect to the social welfare $V(\z^\sigma)-C(\x^\sigma)-\sum_{j\in\J}\underline{q}_j^\sigma$, which can be different from the original social welfare $V(\z)-C(\x)$ in Mechanism 2. Intuitively, we can see that task owners whose task values are less than the minimum payments due to the reserve price, i.e., $u_j \cdot z_j^{\sigma} (\b,\bv )<\underline{q}_j^\sigma$, will not participate in the auction. Hence, the total social welfare is reduced by introducing the reserve price. This completes the proof.
\end{proof}

\end{document}